\documentclass[aip,rsi,amsmath,amssymb,reprint,]{revtex4-1}

\usepackage{graphicx}
\usepackage{dcolumn}
\usepackage{bm}

\begin{document}

\preprint{AIP/123-QED}

\title{Unravelling the contribution of local structures to the anomalies of water: the synergistic action of several factors}

\author{Fausto Martelli}
\affiliation{IBM Research, Hartree Centre, Daresbury, WA4 4AD, United Kingdom}
\email{fausto.martelli@ibm.com}


\begin{abstract}
We investigate the microscopic origin of water's anomalies by inspecting the hydrogen bond network (HBN) and the spatial organization of low-density-liquid (LDL) like and high-density-liquid (HDL) like environments. Specifically, we simulate --via classical molecular dynamics simulations-- the isobaric cooling of a sample composed by 512 water molecules from ambient to deeply undercooled conditions at three pressures, namely 1 bar, 400 bar and 1000 bar. \emph{In correspondence with the Widom line}, (i) the HDL-like dominating cluster undergoes fragmentation caused by the percolation of LDL-like aggregates following a spinodal-like kinetics; (ii) such fragmentation always occurs at a "critical" concentration of $\sim20-30\%$ in LDL; (iii) the HBN within LDL-like environments is characterized by an equal number of pentagonal and hexagonal rings that create a state of maximal frustration between a configuration that promotes crystallization (hexagonal ring) and a configuration that hinders it (pentagonal ring); (iv) the spatial organization of HDL-like environments shows a marked variation. Moreover, the inspection of the global symmetry shows that the intermediate-range order decreases in correspondence with the Widom line, and such decrease become more pronounced upon increasing the pressure, hence supporting the hypothesis of a liquid-liquid critical point. Our results reveal and rationalize the complex microscopic origin of water's anomalies as the cooperative effect of several factors acting synergistically. Beyond implications for water, our findings may be extended to other materials displaying anomalous behaviours.
\end{abstract}

\maketitle

\section{\label{sec:level1}Introduction}
In 1933, Bernal and Fowler provided the first microscopic description of the structure of liquid water and identified the presence of an underlying hydrogen-bond network (HBN) connecting neighbouring molecules in a tetrahedral-like configuration~\cite{bernal_fowler}. 
In order to account for the observed changes of water upon cooling, Bernal and Fowler suggested that water is characterized by a continuous transformation of the HBN from a denser quartz-like structure four-fold coordinated, to a less denser tridymite structure four-fold coordinated also, where the tridymite structure is the structure of hexagonal ice (Ih). Only a few years later, in 1939, Pauling recognized that the HBN is a characteristic that differentiate water from simpler liquids~\cite{pauling}, a notion that is nowadays widely accepted.
In 1951, People further developed the concept of HBN framing it in a random network picture, showing that the HBN can accommodate arrangements in which molecules in the second or third shells of neighbours can collapse towards the first coordination shell of a central molecule~\cite{people}. To further enrich this emerging complex picture, in 1976, Speedy and Angell observed that, at variance with other liquids, the isothermal compressibility of supercooled water increases upon cooling, suggesting the presence of a thermodynamic singularity at -45$^{\circ}$C~\cite{speedy_angell}. This hypothesis was further supported by the evidence that other thermodynamic response functions also tend to diverge upon supercooling. In 1992, based on the results of classical molecular dynamics simulations, Poole \emph{et al} hypothesized that a liquid-liquid critical point (LLCP) located at deeply undercooled conditions and low/intermediate pressures could be the source of the observed thermodynamic anomalies~\cite{poole_nature,poole_pre_1,poole_pre_2}. In correspondence with the LLCP, two liquids (an high-density liquid --HDL-- and a low-density liquid --LDL--) are in metastable equilibrium with each other. The existence of the LLCP has been proven via numerical simulations for a molecular model of water~\cite{mio_nature}, and several other computational~\cite{poole_nature,poole_pre_1,poole_pre_2,xu_pnas,abascal_2010,liu_2010,abascal_2011,corradini_2011,li_2013,palmer_2013,yagasaki_2014,holten_2014,smallenburg_2014,smallenburg_2015,ni_2016,pathak_2016,biddle_2017,palmer_2018} and experimental~\cite{mishima_2002,mishima_1998,mishima_2000,fuentevilla_2006,kanno_2006,mallamace_2008,mishima_2010,bertrand_2011,holten_2012_2,sellberg_2014,sellberg_2014_nature,fanetti_2014_2,kim_2017} investigations point toward the same confirmation. 
A remarkable evidence in favour of the LLCP is the polyamorphic character that water acquires at low temperatures and low/intermediate pressures. Two amorphous states (a low-density amorphous --LDA-- ice, and an high-density amorphous --HDA-- ice) are separated by a first-order phase boundary~\cite{mishima_1985,mishima_1994,mishima_2002,klotz)2005,winkel_2011,martelli_hyperuniformity,martelli_searching} and correspond to the glassy state of LDL and HDL. A very high-density amorphous state has also been hypothesized~\cite{loerting_2001}. Signatures of LDA and HDA have been found in the inherent potential energy surface (IPES, a collection of local potential energy minima resulting from systematic quenches along a liquid trajectory allowing to separate packing from thermal motion effects~\cite{stillinger_1984_1,stillinger_1984_2}) of \emph{ab initio} liquid water at ambient conditions, where the LDA-like and the HDA-like environments are characterized by having distinct HBNs~\cite{santra_2015}. The distinct HBNs in the IPES of \emph{ab initio} liquid water indicate that there is an energy barrier associated with the different HBNs, and this barrier could be large enough at low temperatures and pressures to account for the first-order nature of the LDA-HDA phase boundary. Two distinct local structures have also been observed experimentally in liquid water from ambient to supercooled conditions, and interpreted as LDL-like and HDL-like environments~\cite{torre_2013} as well as under pressure~\cite{fanetti_2014_1}. Water can then be regarded as a dynamical mixture of two different local structures --or two states-- whose fraction changes with the thermodynamic conditions, accounting for the thermodynamic anomalies~\cite{tanaka_2000,tanaka_2000_2,russo_2014,holten_2012,holten_2014,nilsson_2015,biddle_2017,anisimov_2018}. Evidence supporting the existence of two states in liquid water has been found by Raman spectroscopy~\cite{walrafen_1964,walrafen_1967,walrafen_1986}, femtosecond mid-IR pump-probe spectroscopy~\cite{woutersen_1997}, time-resolved optical Kerr effect spectroscopy~\cite{torre_2013}, X-ray absorption~\cite{sellberg_2014} and emission spectroscopies~\cite{huang_2009}. \\
\begin{figure}
  \begin{center}
   \includegraphics[scale=.50]{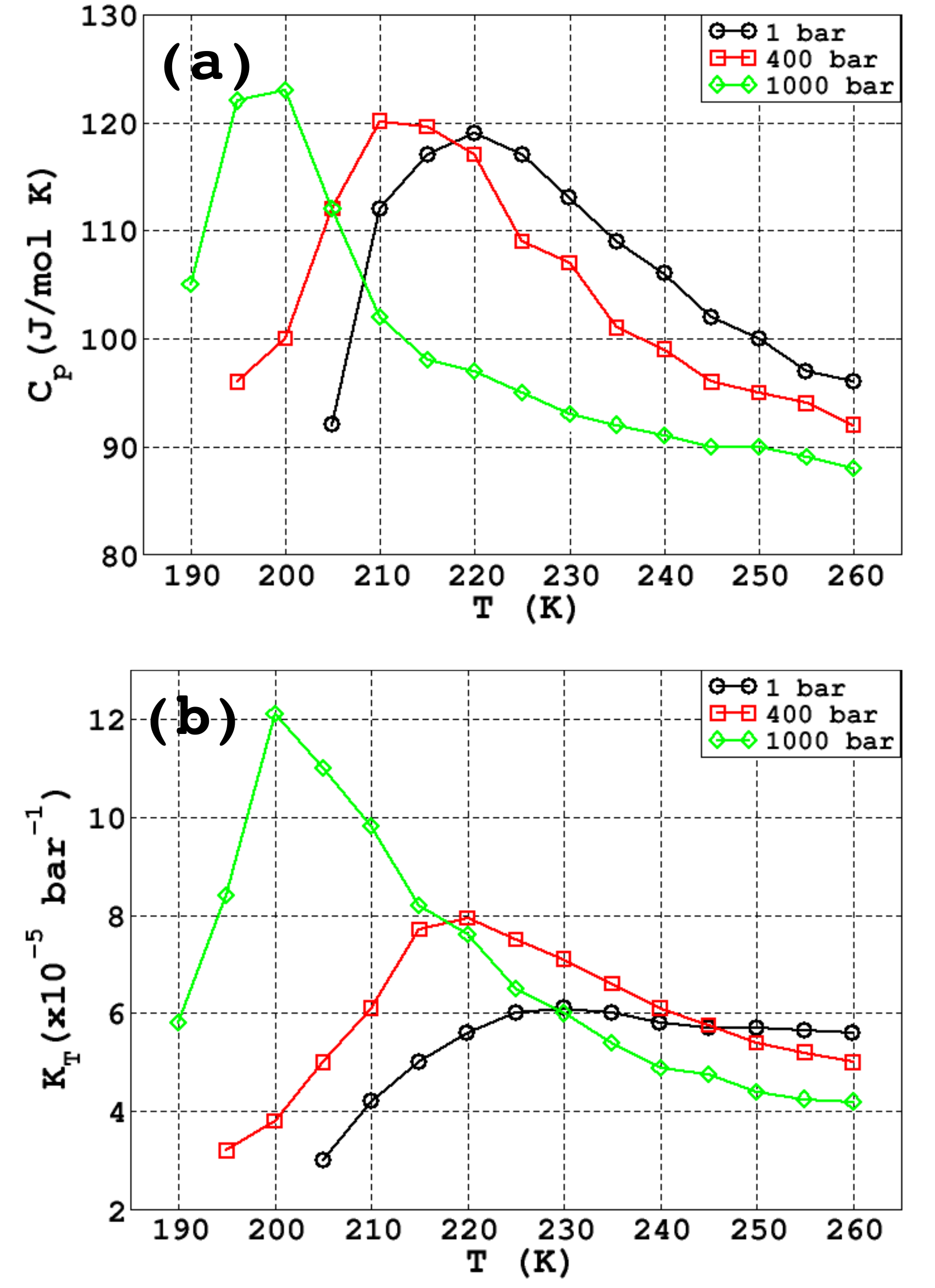}
    \caption{\label{fig:thermo} Temperature dependent isobaric heat capacity (a) and isothermal compressibility (b) at 1 bar, 400 bar and 1000 bar.}
  \end{center}
\end{figure}
Within the LLCP scenario, a line of maxima of the correlation length emanates from the LLCP, at which the correlation lengths diverge. This line, known as the Widom line (WL), is the locus of maxima of the thermodynamic response functions, and where a strong to fragile crossover occurs~\cite{starr_1999,faraone_2004,xu_pnas,liu_2005,mallamace_2006,zhang_2009,gallo_2010,gallo_2012,gallo_2012_2,wang_2015,YXu_2016,demarzio_2016,demarzio_2017}, suggesting that thermodynamic anomalies are intimately linked with dynamic anomalies. In this work, we define T$_{WL}$ the temperature range in which, at a given pressure, the isobaric heat capacity $C_P$ and the isothermal compressibility $\kappa_T$ show a maximum. In fig.~\ref{fig:thermo} we report $C_P$ (panel (a)) and $\kappa_T$ (panel (b)) for the three pressures here investigated. We can observe that T$_{WL}$ is $\sim220-230$K for 1 bar, $\sim210-220$K for 400 bar and $\sim200-205$ for 1000 bar. In water, the WL is negatively sloped, so the Clapeyron equation implies that the high-temperature phase is HDL dominated, and the low-temperature phase is LDL dominated. Recently, Shi \emph{et al} provided a microscopic basis for a two-state description of water, and proposed that the emergent fragile-to strong transition is actually a crossover between two Arrhenius regimes with different activation energies, bringing into question the glass-transition scenario~\cite{shi_2018_1,shi_2018_2}. \\
The outstanding importance of water anomalies is nowadays recognized to play a central role in processes in several fields, from physics and chemistry, to biology, material science, geology and climate modeling~\cite{kauzmann,angell_1995,mishima_1998,debenedetti_2003}. For this reason, a lot of effort has been devoted to rationalizing water's anomalous behaviours, but a coherent and simple picture has yet to be found.\par
In this article, we have performed equilibrium classical molecular dynamics (MD) simulations crossing the WL at three pressures, namely 1 bar, 400 bar and 1000 bar. Therefore, we concern ourselves only with the lines of maxima determined along isobars for the remainder of the current work\footnote{The Widom line can also be defined, e.g., on the isothermal pathway~\cite{fomin_2015,schienbein_2018}}. We have observed an hitherto unknown intimate connection between the percentage of LDL-like molecules and their HBN, and between HDL-like molecules and their spatial organization occurring in correspondence with the WL. Our results indicate that water's anomalies occur under specific conditions of composition and connectivity, and enlighten the role of the underlying HBN. Therefore, our results elucidate the basic microscopic physical picture hidden behind water's anomalies, and may represent a generic feature of anomalous liquids at large.\par
The article is organized as follows: in Section~\ref{sims} we describes the molecular dynamics set up and two order parameters adopted in the present work. In Section~\ref{results} we present our results, while final remarks are reported in Section~\ref{conclusions}.

\section{Simulation details}\label{sims}
In this Section we describe the protocol we have implemented in our simulations and the two order parameters we have employed to investigate the temperature-dependence of the global order while crossing the WL and to distinguish local environments.
\subsection{Classical molecular dynamics simulations}
We have performed classical MD simulations of a sample of 512 water molecules interacting with the TIP4P/2005 potential~\cite{tip4p2005} in the NPT ensemble. We have employed Nos\'e-Hoover thermostat~\cite{nose,hoover} with 0.2 ps relaxation time to maintain constant temperature, and Parrinello-Rahman barostat~\cite{parrinello_rahman} with 2 ps relaxation time to maintain constant pressure. We have explored three pressures, namely 1 bar, 400 bar and 1000 bar, and isobarically scanned temperatures from 300 K down to 205-190 K depending on the pressure. We have performed simulations with the GROMACS 5.0.1 package~\cite{gromacs}. We have truncated short-range interactions at 9.5 \AA, and  we have computed long range electrostatic terms using particle mesh Ewald with a grid spacing of 1.2 \AA. At each state point, we have computed and carefully monitored the decay of the self-part of the intermediate scattering function (ISF) with time~\cite{hansen}. All reported trajectories are at least 500 times longer that the structural relaxation time as computed from the ISF. Depending on the thermodynamic conditions, production runs vary between 50 ns and 12 $\mu$s. No sign of crystallization have been observed. 
\subsection{The order parameters}
In this section we describe the two order parameters employed in this work.
\subsubsection{The score function}
The local environment of an atomic site $j$ in a snapshot of a molecular dynamics or Monte Carlo simulation defines a local pattern formed by $M$ neighboring sites. Typically these include the first and/or the second neighbors of the site $j$. There are $N$ local patterns, one for each atomic site $j$ in the system. The local reference structure is the set of the same $M$ neighboring sites in an ideal lattice of choice, the spatial scale of which is fixed by setting its nearest neighbor distance equal to $d$, the average equilibrium value in the system of interest.
For a given orientation of the reference structure and a given permutation $\mathcal{P}$ of the pattern indices, we define the LOM $S(j)$ as the maximum overlap between pattern and reference structure in the $j$ neighborhood by:
\begin{equation}
  S(j)=\max_{\theta,\phi,\psi;\mathcal{P}}\left\{\prod_{i=1}^{M}\exp\left(-\frac{\left| \mathbf{P}^{j}_{i\mathcal{P}}-\mathbf{A}^{j}\mathbf{R}_{i}^{j}\right|^2}{2\sigma^{2}M}\right)\right\}
  \label{eq:Eq1}
\end{equation}
Where $\theta, \phi$ and $\psi$ are Euler angles, $\mathbf{P}^{j}_{i\mathcal{P}}$ and $\mathbf{R}_{i}^{j}$ are the pattern and the reference position vectors in the laboratory frame of the $M$ neighbors of site $j$, respectively, and $\mathbf{A}^{j}$ is an arbitrary rotation matrix about the pattern centroid. The parameter $\sigma$ controls the spread of the Gaussian functions ($\sigma=d/4$ in this work, where $d$ is the characteristic length of the local pattern). The LOM satisfies the inequalities $0 \lesssim S(j) \leq 1$. The two limits correspond, respectively, to a local pattern with randomly distributed points ($S(j)\rightarrow 0$) and to an ordered local pattern matching perfectly the reference ($S(j)\rightarrow 1$). We also define a global order parameters based on $S(j)$, as the average score function $S$:
\begin{equation}
  S=\frac{1}{N}\sum_{j=1}^{N}S(j)
  \label{eq:Eq2}
\end{equation}
We have recently employed the LOM and the score function to enlighten structural properties of water and its homogeneous crystallization at various conditions~\cite{martelli_LOM,martelli_searching,martelli_fop,samatas_2018}, and -as a collective variable- to drive the formation of boron-nitride nanotubes~\cite{santra_bnnt}. The details of the numerical algorithm can be found in Ref~\cite{martelli_LOM}.\\
In this work, we have employed eq.~\ref{eq:Eq2} to measure the global symmetry of our samples upon isobarically cooling and crossing the Widom line at the three pressures here investigated.

\subsubsection{The local structure index}
We order the set of radial oxygen-oxygen distances ${r_j}$ corresponding to the $N$ neighbouring molecules that are within a cut-off distance of $3.7$ \AA from the reference molecule as follows: $r_1<r_2<...<r_j<r_{j+1}...<r_N<3.7<r_{N+1}$. The local structure index (LSI) value $I$ is then defined as the inhomogeneity in this distribution of radial distances, i.e.,
\begin{equation}
I(i)=\frac{1}{n(i)}\sum_{j=1}^{n(i)}\left[\Delta(j;i)-\bar{\Delta}(i)\right]^2
\label{eq:Eq3}
\end{equation}
where $\Delta(j;i)=r_{j+1}-r_{j}$ and $\bar{\Delta}(i)$ is the average over all neighbours $j$ of a molecule $i$ within a given cutoff. Hence, $I$ provides a convenient quantitative measure of the fluctuations in the distance distribution surrounding a given water molecule within a sphere defined by a radius of $\sim3.7$ \AA~\cite{LSI_1,LSI_2}. In doing so, the index $I$ measures the extent to which a given water molecule is surrounded by well-defined first and second coordination shells.\\
In this work, we employ the LSI to characterize local environments in our samples as a function of the temperature and of the pressure.

\section{Results}\label{results}
In this section we present and discuss the main results of our work. Computational and experimental studies on the structural changes of liquid water have shown that the short range order (SRO) is only marginally affected by the change in pressure and temperature~\cite{soper_2000,svishchev_1993,schwegler_2000,saitta_2003,sciortino_1990,sciortino_1991,kumar_2006,franzese_2007,lapini_2016}. Rather, relevant effects occur beyond the SRO, at the level of the intermediate range order (IRO)~\cite{martelli_LOM}. In particular, in 2000, Soper and Ricci showed that part of the second shell collapses towards the first shell increasing the local density and pointing out the importance of the second shell~\cite{soper_2000}. The role of the second shell has consequently been enlightened via several local descriptors such as, e.g., the distance of the fifth neighbour $g_5$~\cite{cuthbertson_2011}, the LOM described in eq.~\ref{eq:Eq1}~\cite{martelli_LOM}, and the LSI~\cite{LSI_1,LSI_2} that measures the degree of inhomogeneity between the first and the second shell. Wiktfeldt \emph{et al} showed that, at the level of the IPES, the LSI distribution in liquid water has a bimodal distribution that, in correspondence with the WL, gives a 1:1 ratio between HDL-like and LDL-like environments~\cite{wikfeldt_2011}. It is worthy to remark that, with respect to Wiktfeldt \emph{et al}, this work inspects the behaviour of water in the presence of thermal noise. Consequently, Russo and Tanaka recognized the existence of locally favoured structures by measuring the degree of translational order on the second shell, enlightening their role in the anomalies of water~\cite{russo_2014}.
\subsection{Structural order}
\subsubsection{Global order from the score function}
\begin{figure}
  \begin{center}
   \includegraphics[scale=.23]{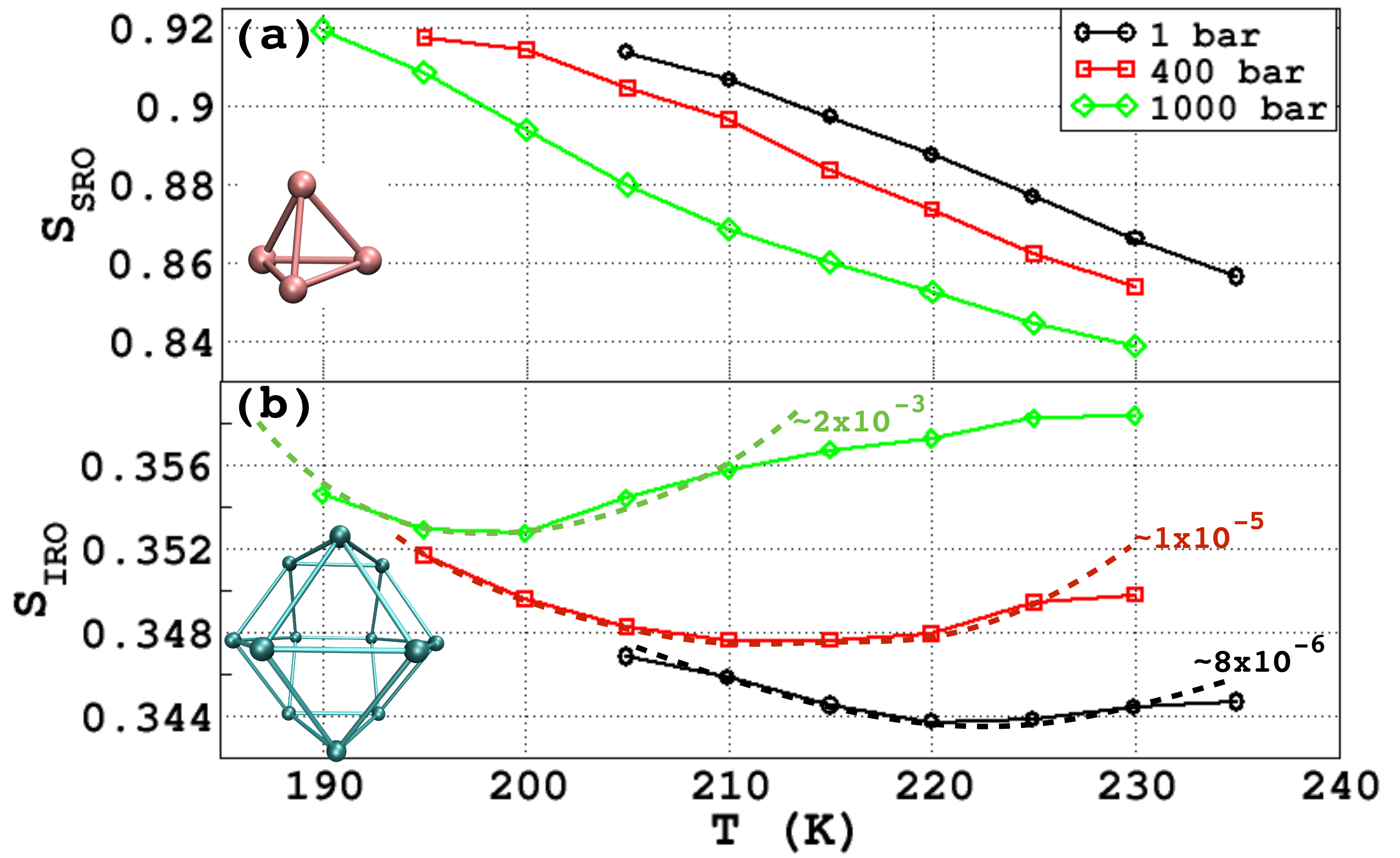}
    \caption{\label{fig:scores} (a) Short range order using, as a reference, a perfect tetrahedron, for the cooling at 1 bar (black circles), at 400 bar (red squares) and at 1000 bar (green diamonds), respectively. (b) Intermediate range order using, as a reference, the anticuboctahedron. Dashed lines correspond to the quadratic fittings.}
  \end{center}
\end{figure}

In this paragraph, we inspect the local order of liquid water by separating the contribution of the first shell from the contribution of the second shell employing the score function described above. In Fig.~\ref{fig:scores} (a) we report $S_{\text{SRO}}$, the score function (eq.~\ref{eq:Eq2}) for the SRO measured using, as a reference, a perfect tetrahedron. As expected, for a given temperature the $S_{\text{SRO}}$ decreases upon increasing the pressure, while it increases upon fixing the pressure and isobarically cooling. The increment of the $S_{\text{SRO}}$ upon isobarically cooling the sample is an effect caused by the reduction of the thermal energy, as also measured by the increase of the intensity of the first peak in the oxygen-oxygen radial distribution function. \\
In Fig.~\ref{fig:scores} (b) we report $S_{\text{IRO}}$, the score function (eq.~\ref{eq:Eq2}) for the IRO computed using, as a reference, the anticuboctahedron that describes the spatial configuration of the oxygen atoms at the level of the second shell in Ih. With respect to $S_{\text{SRO}}$, the $S_{\text{IRO}}$ decreases upon increasing the pressure at a given temperature. Since in our maximization algorithm we let the reference structure to rescale its dimensions with respect to the local pattern~\cite{martelli_LOM}, the increase of $S_{\text{IRO}}$ upon increasing the pressure indicates that the second shell tends to increase the local packing~\cite{martelli_searching}. Remarkably, upon cooling isobarically our samples, the $S_{\text{IRO}}$ shows a minimum in correspondence with the location of the WL. We have fitted the loci of each minimum along with the two closest state points on the left and on the right, with a quadratic function at all pressures (dashed lines in fig.~\ref{fig:scores} (b)). Notably, the amplitude of the quadratic function shrinks from $\sim 8\times 10^{-6}$ at 1 bar, to $\sim 10^{-5}$ at 400 bar and $\sim 2\times10^{-3}$ at 1000 bar. The minima in $S_{\text{IRO}}$ is indicative of the decrease in the local order caused by the density fluctuations that become more pronounced approaching the hypothesized LLCP~\cite{abascal_2010}. Therefore, the shrinkage of the amplitude supports the LLCP scenario for this water model and confirm that the thermodynamic and dynamic water anomalies have a structural counterpart.

\subsubsection{High-density and low-density environments from the local structure index}
In order to enlighten the role played by local environments in the emergence of the anomalies of water, we have characterized LDL-like and HDL-like molecules based on the corresponding value of the LSI. As reported by Wikfeldt \emph{et al}~\cite{wikfeldt_2011} and  Appignanesi \emph{et al}~\cite{appignanesi_2009,montes_de_oca}, the LSI distribution in the IPES is a bimodal function with an isosbestic point $I_{is}$ mostly unaffected by the thermodynamic conditions and located at $I_{is}\sim0.13$ \AA$^2$. Thus, it provides a well defined criteria to distinguish LDL-like and HDL-like environments at supercooled conditions, as recently shown by Shi and Tanaka~\cite{shi_2018_3}. In this work, since water's anomalies occur in the presence of thermal energy, we have borrowed the $I_{is}$ from the IPES to define LDL-like and HDL-like environments at the level of the dynamical trajectories, i.e., in the presence of thermal energy~\footnote{The position of the isosbestic point in the presence of thermal energy fluctuates within the range $0.12-0.14$ \AA$^2$. Any change within this range would produce very minor quantitative changes in our results. The position of the isosbestic point also depends on the cutoff employed in the evaluation of eq.~\ref{eq:Eq3}~\cite{accordino_2011}}. 
\begin{figure}
  \begin{center}
   \includegraphics[scale=.33]{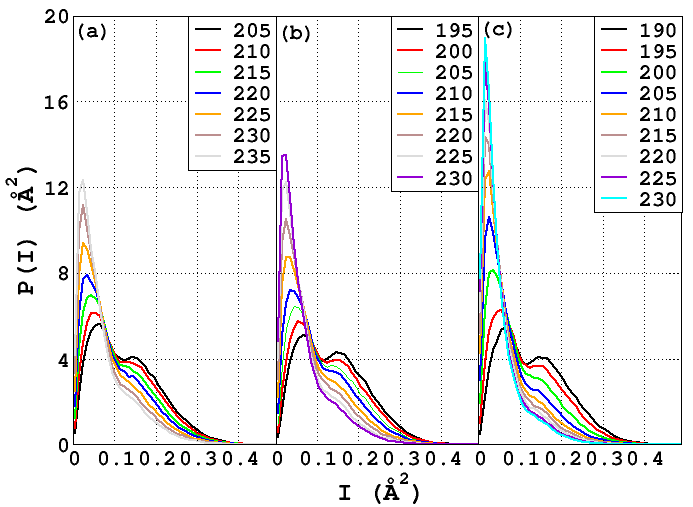}
    \caption{\label{fig:PLSI} PLSI distributions computed at different temperatures and pressures: (a) for the case of 1 bar, (b) for the case of 400 bar and (c) for the case of 1000 bar.}
  \end{center}
\end{figure}
The advantages of investigating local structures in the presence of thermal energy have been also recently enlightened by Russo and Tanaka~\cite{russo_2014} and by Saito \emph{et al}~\cite{saito_2018}. \par
In fig.~\ref{fig:PLSI} we report the probability density distribution of the LSI order parameter ($P(I)$) and we begin our discussion by analyzing how $P(I)$ depends on the thermodynamic conditions. Panel (a) reports $P(I)$ computed at 1 bar in the temperature range $205\leq T\leq 235$K, panel (b) at 400 bar and in the temperature range $195\leq T\leq 230$K and panel (c) at 1000 bar and in the temperature range $190\leq T\leq230$K. The bimodal character of $P(I)$ become more pronounced upon cooling the samples. This finding is indicative of the fact that, at high temperatures, the majority of the molecules are situated in HDL-like locally disordered environments. Upon cooling isobarically, we observe a systematic decrease in the relative population of water molecules with lower LSI values (i.e., molecules in locally disordered environments) coupled with an increase in the relative population of water molecules with higher LSI values (i.e., molecules in locally ordered environments). In the remainder of this work we will focus only on the aforementioned temperature intervals.\\
\begin{figure}
  \begin{center}
   \includegraphics[scale=.23]{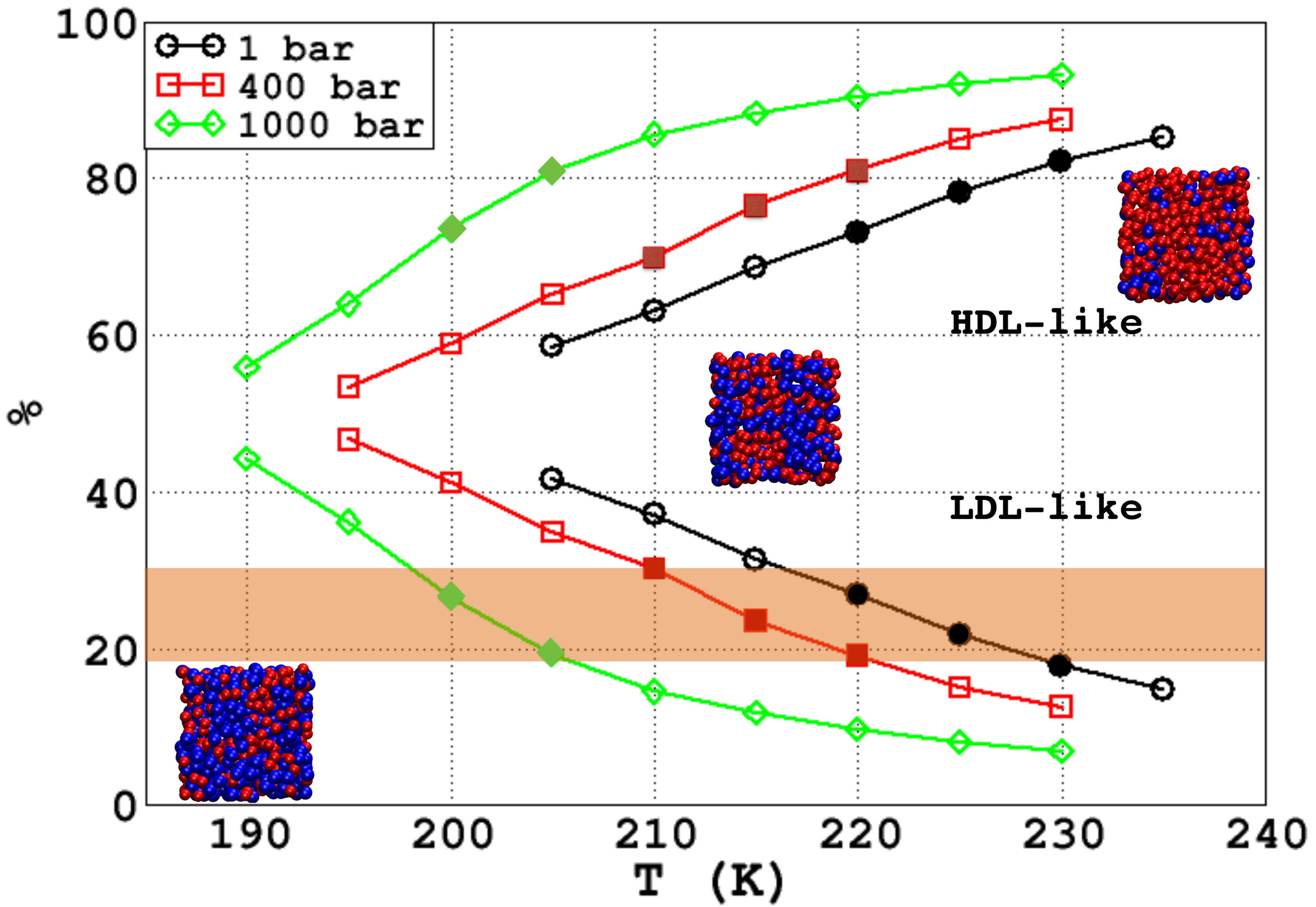}
    \caption{\label{fig:LDL} Percentage of HDL-like and of LDL-like molecules computed from the LSI (see main text for the definition) for the cooling at 1 bar (black circles), at 400 bar (red squares) and at 1000 bar (green diamonds), respectively. Filled symbols represent the state points at the corresponding Widom lines. The colored stripe emphasize the critical concentration. Three representative snapshots describe local environments at 400 bar and 230K, 215K and 195K. Red spheres represent HDL-like environments, while blue spheres represent LDL-like environments.}
  \end{center}
\end{figure}
In fig.~\ref{fig:LDL} we report the percentage of LDL-like molecules and of HDL-like molecules as a function of the temperature for the three pressures here investigated. As expected, the percentage of LDL-like environments increases upon cooling the samples at all pressures, indicating that the space between the first and the second shell of neighbours become less and less populated and the local environments progressively more tetrahedral, as a result of the reduction of thermal energy. This trend can also be visualized graphically by observing the three representative snapshots respectively obtained at $230$, $215$ and $195$ K and at 400 bar (similar pictures hold for the other pressures inspected in this work). Red spheres represent HDL-like molecules, while blue spheres represent LDL-like molecules. Remarkably, in correspondence with the WL (filled simbols), the composition of the sample relative to LDL acquire the same value in the range of $\sim20-30\%$ for \emph{all} pressures (colored stripe). Since this behaviour occur at the same concentrations of LDL-like particles at all pressures, we infer that this concentration represents a kind of "critical mass" needed to sustain changes in the dynamical and thermodynamic behaviour. We therefore infer that this relatively narrow range of composition could be a key ingredient for a system like water to display anomalous behaviours, and a common feature for other liquids also displaying anomalous behaviours. We will provide further proofs of the "criticality" of such concentration in the remainder of the article.
\subsection{Clustering of LDL-like and of HDL-like environments}
In order to delve deeper into the role of LDL-like and HDL-like environments in the emergence of the anomalies of water, we have inspected the tendency of LDL-like and of HDL-like environments to aggregate and form clusters. We here define a cluster an aggregate of at least two water molecules belonging to the same class -LDL or HDL- as measured by the LSI, and separated by an oxygen-oxygen distance shorter than $3.3$ \AA.\par
\begin{figure}
  \begin{center}
   \includegraphics[scale=.33]{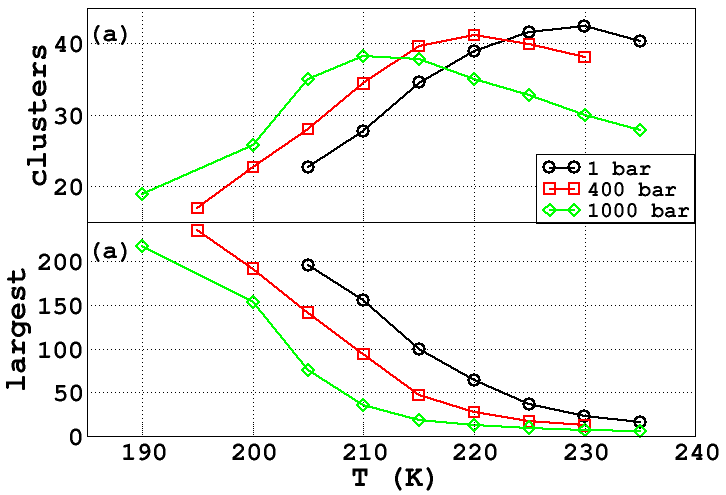}
    \caption{\label{fig:clust_LDL} (a) Number of LDL-like clusters. (b) Size of the largest LDL-like cluster}
  \end{center}
\end{figure}
\begin{figure}
  \begin{center}
   \includegraphics[scale=.33]{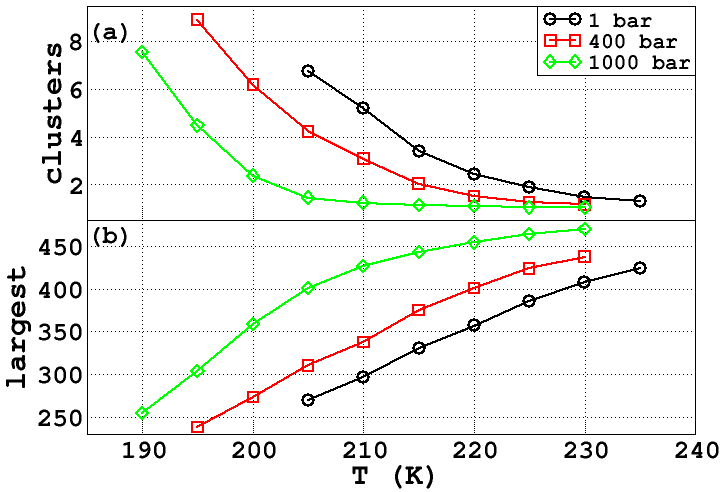}
    \caption{\label{fig:clust_HDL} (a) Number of HDL-like clusters. (b) Size of the largest HDL-like cluster}
  \end{center}
\end{figure}
\begin{figure}
  \begin{center}
   \includegraphics[scale=.23]{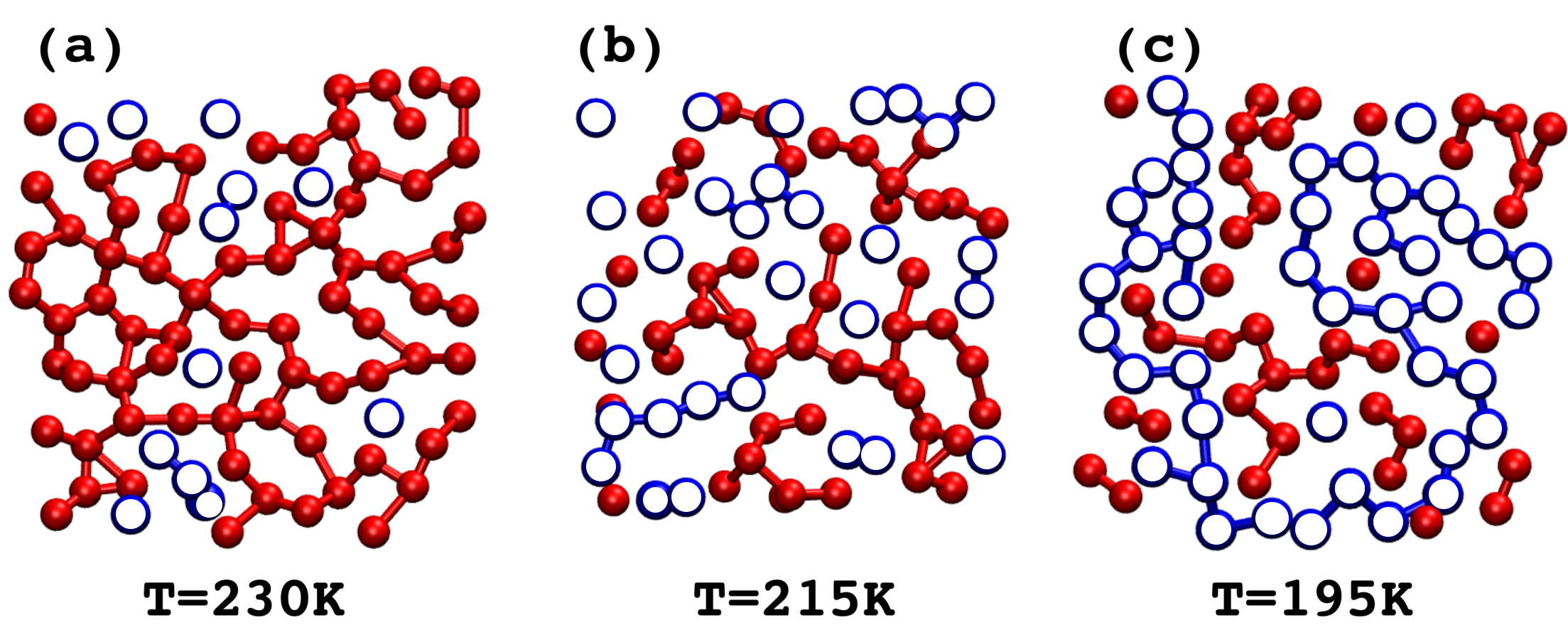}
    \caption{\label{fig:proj} Two dimensional slices of the LDL-like network (blue, open spheres) and of the HDL-like network (red, filled spheres) at 400 bar and three temperatures: T=230K (a), T=215K (b) and T=195K (c).}
  \end{center}
\end{figure}

The number of LDL-like clusters is reported in fig.~\ref{fig:clust_LDL} (a) and shows a non monotonic behaviour with maxima at temperatures slightly higher than the WL and a decreasing trend starting, upon cooling, in correspondence with the T$_{WL}$. This observation indicates that, in correspondence with the T$_{WL}$, the LDL-like clusters start forming an extended network further expanding upon cooling. As a confirmation, in fig.~\ref{fig:clust_LDL} (b) we report the size of the largest LDL-like cluster. We can observe that the pace of growth of the largest cluster shows two distinct regimes: a slow growth at temperatures above T$_{WL}$, ad a faster growth starting in correspondence with T$_{WL}$ upon cooling the sample. This trend is indicative of the fact that smaller clusters are absorbed into the largest expanding cluster. At high temperatures, the low number of LDL-like clusters (fig.~\ref{fig:clust_LDL} (a)) and the low percentage of LDL-like molecules (fig.~\ref{fig:LDL}) indicate that the sample is composed by many LDL-like small, scattered clusters. On the other hand, in correspondence with the WL, the percentage of LDL-like molecules reaches the critical concentration of $\sim20-30\%$, high enough for LDL-like clusters to percolate and form a large aggregate, further proving that this amount of LDL-like molecules represent a critical concentration. \\
In fig.~\ref{fig:clust_HDL} (a) we report the number of HDL-like clusters, while in fig.~\ref{fig:clust_HDL} (b) we report the size of the largest HDL-like cluster. At temperatures above the T$_{WL}$, the system is composed by one large HDL-like aggregate spanning the entire simulation box and enclosing smaller LDL-like clusters. Upon cooling, the largest HDL-like cluster gets continuously depleted (panel (b)) at the expenses of LDL-like environments (fig.~\ref{fig:LDL}). Remarkably, the largest HDL-like cluster gets scattered into 2-3 smaller clusters in correspondence with the T$_{WL}$ (panel (a)), confirming that LDL-like clusters percolate inside the HDL-like large cluster, fragmenting it into smaller domains. Therefore, the term "critical" referred to the $\sim20-30\%$ amount of LDL-like molecules marks the limiting value of LDL-like environments at which these molecules form an aggregate large enough to percolate within the HDL-like network and to split it. Upon further cooling isobarically the samples, the number of HDL-like clusters increases (panel (a)) while the dimension of the HDL-like cluster keep decreasing (panel (b)), indicating that the LDL-like network is now further percolating inside the HDL-like network, creating several HDL-like patches. \\
In summary, by inspecting the evolution of LDL-like and of HDL-like clusters as a function of the temperature and of the pressure, we observe that the system is composed by a large HDL-like cluster at temperatures above T$_{WL}$. Upon cooling, the percentage of LDL-like molecules increases and, in correspondence with T$_{WL}$, it reaches a "critical" concentration of $\sim20-30\%$ that allows the LDL-like clusters to connect each other, percolating and fragmenting the large HDL-like cluster. Upon further cooling isobarically the sample, the LDL-like network keep expanding generating a fast-growing large LDL-like cluster. This picture is consistent with a spinodal-like decomposition, in which phase separation occur in correspondence with T$_{WL}$. We have recently observed a similar behaviour during the simulated isothermal compression of LDA to produce HDA via a mild first-order phase transition~\cite{martelli_searching}.\\ 
In fig.~\ref{fig:proj} we report a two-dimensional 4 \AA-thick slice of our sample at 400 bar and at three temperatures: T=230K --(a)--, T$\sim$T$_{WL}$=215K --(b)-- and T=195K --(c)--. Blue spheres represent LDL-like environments, while red spheres represent HDL-like environments. At T=230K, the sample is permeated by the large HDL-like cluster and its underlying HBN (red sticks), embracing LDL-like molecules and clusters. Upon cooling the sample to T$\sim$T$_{WL}$=215K, the main HDL-like network no longer span the entire simulation box; rather, it is now fragmented and alternated with the likewise fragmented LDL-like  network (blue sticks). Finally, at T=195K we can observe that the HDL-like and the LDL-like networks are now almost independent and separated. The formation of two almost independent networks at low temperatures, allows us to speculate that the tendency of the system is to effectively unmix and separate the two HBNs, a mechanism that occur in its entirety in correspondence with the LLCP. Moreover, since local environments characterized by different structural properties diffuse with different velocities~\cite{xu_2009,mallamace_2013}, the drastic changes in terms of composition, clustering and cluster growth occurring in correspondence with T$_{WL}$ may explain the fragile-to-strong crossover occurring in correspondence with the T$_{WL}$.
\subsection{Radial distribution functions and the hydrogen bond network}
\begin{figure}
  \begin{center}
   \includegraphics[scale=.32]{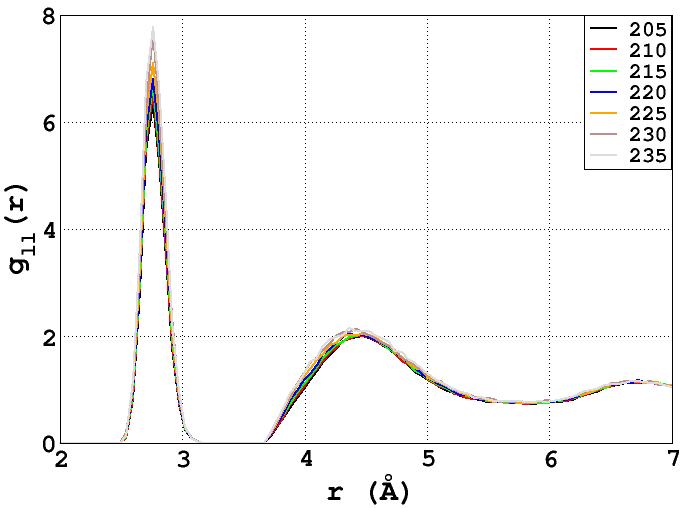}
    \caption{\label{fig:rdf_ll} Radial distribution function computed between LDL-like oxygens at 1 bar in the temperature range $205\lesssim T\lesssim235$K.}
  \end{center}
\end{figure}

In the previous paragraph we have observed that the clustering of HDL-like and LDL-like networks plays a fundamental role in the anomalies of water in correspondence with the WL. We now investigate the details of the spatial organization and of the HBN for both LDL-like and HDL-like environments. The definition of HB follows Ref.~\cite{chandler_HB}. In this regard, any quantitative measure of HBs in liquid water is somewhat ambiguous, since the notion of a HB itself is not uniquely defined. However, qualitative agreement between many proposed definitions have been deemed satisfactory over a wide range of thermodynamic conditions~\cite{prada_2013,shi_2018_2}.\\ 
We have inspected the spatial correlation via the two-body radial distribution functions (RDFs) computed \emph{among} the LDL-like and the HDL-like water molecules only, and we have inspected the topology of the HBN using the ring statistics analysis, a tool which has been instrumental in theoretically characterizing the amorphous states LDA and HDA~\cite{martonak_2004,martonak_2005}, and the HBN in the IPES of \emph{ab initio} liquid water at ambient conditions~\cite{santra_2015}. \\
In fig.~\ref{fig:rdf_ll} we report g$_{ll}$(r), the RDFs computed among LDL-like oxygens only at 1 bar, in the temperature range $205\lesssim T\lesssim 235$K. Similar distributions characterize higher pressures. We can observe that the first and the second peak are very well separated, indicating that the space between the first and the second shells is mostly empty, hence reflecting the high tetrahedrality of LDL-like environments. Overall, the RDFs are qualitatively similar to the experimental RDF for LDA obtained using the empirical potential structure refinement of neutron diffraction data reported in Ref.~\cite{finney_2002}.\par
\begin{figure}
  \begin{center}
   \includegraphics[scale=.32]{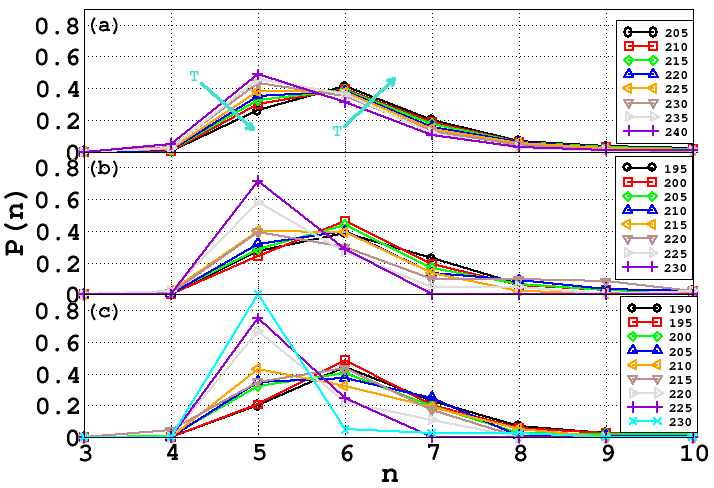}
    \caption{\label{fig:rings_ldl} Probability distributions of the hydrogen-bonded n-folded rings, P(n), for LDL-like molecules, at 1 bar (a), 400 bar (b) and 1000 bar (c). All P(n) have been normalized to unity and therefore do not reflect the total number of rings of a given size.}
  \end{center}
\end{figure}

\begin{figure}
  \begin{center}
   \includegraphics[scale=.25]{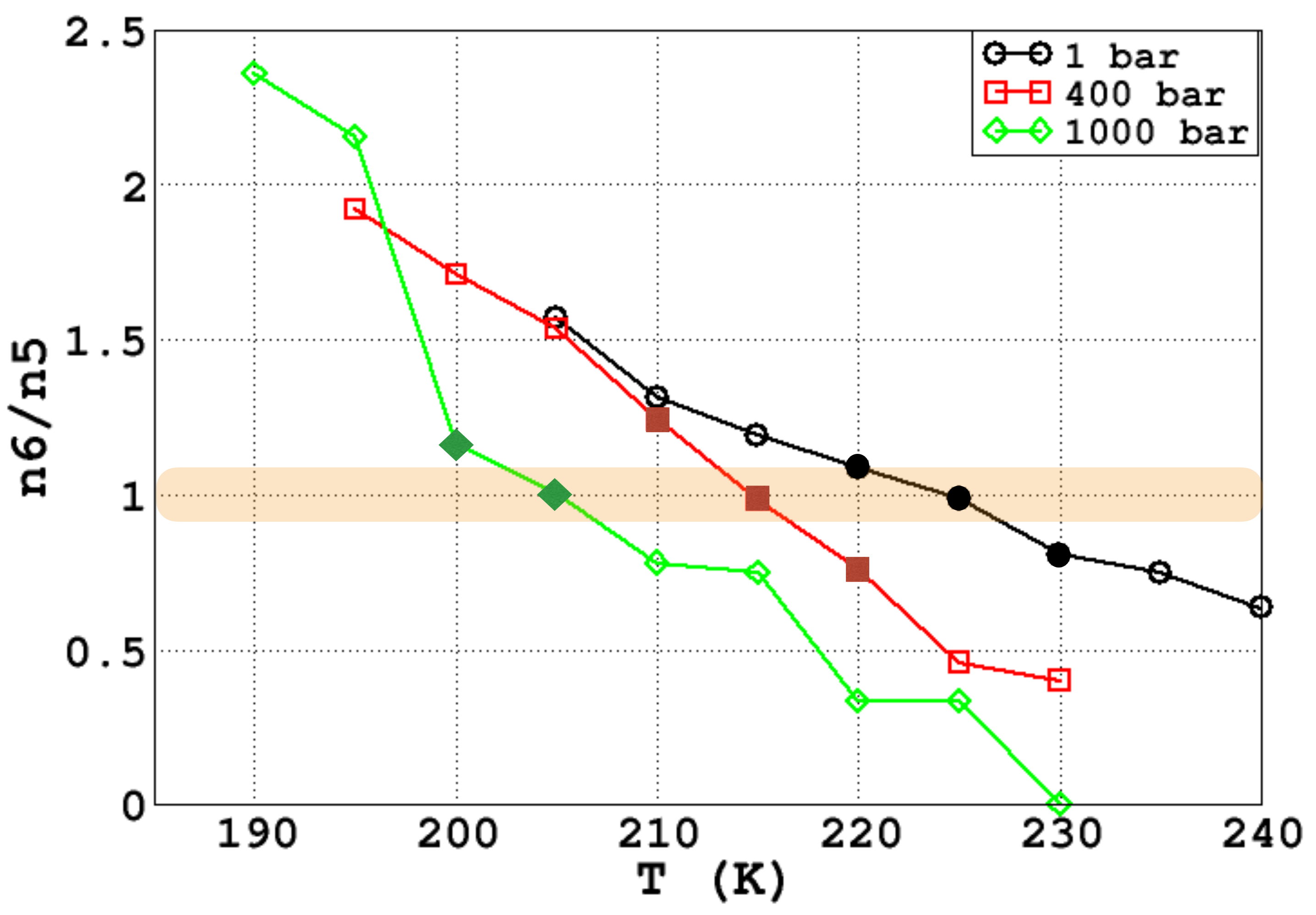}
    \caption{\label{fig:65} Temperature dependence of the ratio between hexagonal (n6) and pentagonal (n5) rings for the three pressures inspected in this work. The colored line emphasizes the inflection point and filled symbols mark the values at the corresponding T$_{WL}$.}
  \end{center}
\end{figure}

We now inspect the topology of the HBN connecting LDL-like molecules. In fig.~\ref{fig:rings_ldl} we report P(n), the normalized probability of having an n-folded ring, n$\in[3,10]$, for LDL-like environments only. Panel (a) shows the distributions at all temperatures at 1 bar, panel (b) the distributions at 400 bar and panel (c) the distributions at 1000 bar. Interestingly, at high temperatures LDL-like molecules are mostly connected via pentagonal rings and the P(n)'s show a maximum in correspondence with n5. We have encountered a similar distribution in the IPES of \emph{ab initio} liquid water at ambient conditions~\cite{santra_2015} and we have concluded that such distribution, that describes a network deviating from the hexagonal one characteristic of the amorphous state LDA, is an artifact due to the quenching at the level of the IPES. On the other hand, our results here indicate that a distribution maximized at n5 is a genuine effect, and we provide here a justification supporting this evidence: at high temperatures, LDL-like environments are low in number and they tend to form small clusters rather than arranging in open wires. Therefore, the pentagon is the spatial arrangement with hydrogen bond distances and angular strains the closest to the hexagonal geometry. Hence, the pentagon represents the most energetically favourable configuration at high temperatures. Upon cooling isobarically the samples, P(n) shifts from a distribution maximized at n5 to a distribution maximized at n6. This observation indicates that LDL-like environments can develop the expected hexagonal network characteristic of the amorphous state LDA only when their number overcome the critical concentration of $\sim20-30\%$. This observation further stresses the importance of such critical concentration. It is worthy to remark that, although the g$_{ll}$(r) shown in fig.~\ref{fig:rdf_ll} are mostly independent with respect to the thermodynamic conditions and resemble --at least qualitatively-- the experimental RDF of LDA, the underlying HBNs are drastically different not only with respect to LDA, but also upon minor changes in the temperature in the liquid phase. It has been recently shown that, upon rapid quenching equilibrated liquid water at different conditions, one obtains LDAs that are indistinguishable in terms of RDFs, but that are characterized by different potential energy landscapes (PELs)~\cite{giovambattista_2016,handle_2018}. Our results suggest that the different underlying HBNs within LDL-like environments in the equilibrated liquid phases may represent the source of the different PELs in the amorphous states.
Remarkably, in correspondence with the WL, P(n) for the three pressures here investigated show an inflection point at which n5=n6. This inflection point represents a state of maximal frustration in a system composed by an equal number of hexagonal rings, known to be the precursor of crystallization, and pentagonal rings, known to frustrate against crystallization. Therefore, we infer that the inflection point at which n5=n6 in the LDL-like HBN represents another key feature, essential for the bulk system to display anomalous behaviours, and may be a general feature of liquids endowed with a LLCP displaying anomalous behaviours.\\
The inflection point is emphasized by the orange line in fig.~\ref{fig:65}, where we report the ratio between the number of hexagonal and pentagonal rings for the three pressures here examined in the temperature range $190\lesssim T\lesssim 240$K. It is possible to observe that the inflection point occurs in correspondence with the T$_{WL}$ (filled symbols) at all pressures.\par
\begin{figure}
  \begin{center}
   \includegraphics[scale=.32]{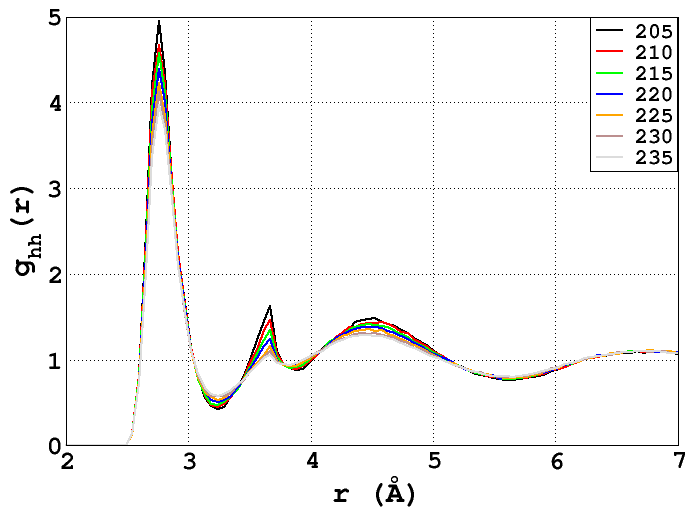}
    \caption{\label{fig:ghh} Radial distribution functions computed between HDL-like oxygens at 1 bar in the temperature range $205\lesssim T\lesssim235$K.}
  \end{center}
\end{figure}

\begin{figure}
  \begin{center}
   \includegraphics[scale=.30]{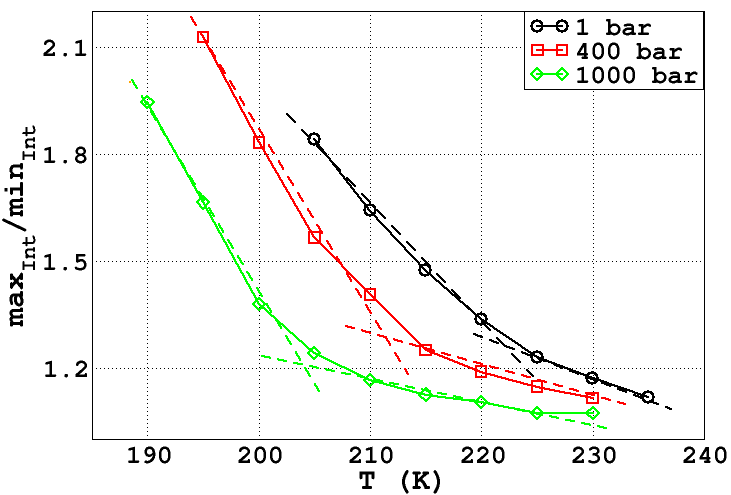}
    \caption{\label{fig:maxmin} Ratio between max$_{\textit{int}}$ and min$_{\textit{int}}$ as a function of the temperature for the three pressures here inspected. Dashed lines serve as guide to the eyes.}
  \end{center}
\end{figure}

\begin{figure}
  \begin{center}
   \includegraphics[scale=.32]{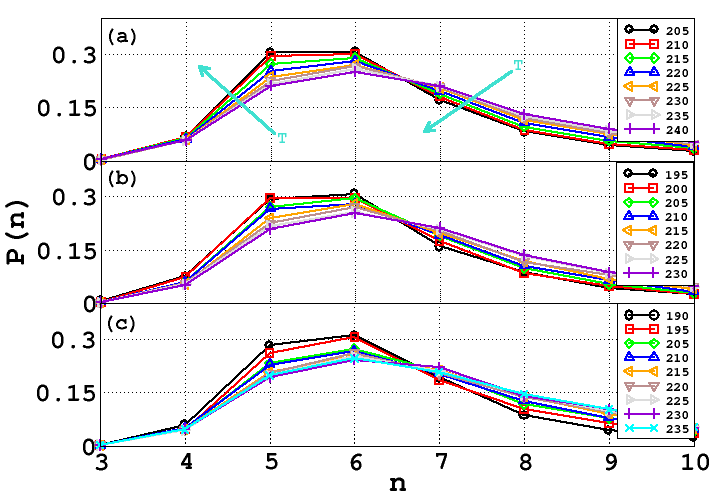}
    \caption{\label{fig:rings_hdl} Probability distributions of the hydrogen-bonded n-folded rings, P(n), for HDL-like molecules, at 1 bar (a), 400 bar (b) and 1000 bar (c). All P(n) have been normalized to unity and therefore do not reflect the total number of rings of a given size.}
  \end{center}
\end{figure}

We now focus our attention on the HDL-like environments. At variance with the LDL-like environments, the g$_{hh}$(r) show a marked temperature dependence that propagates beyond the SRO (fig.~\ref{fig:ghh}). In fig.~\ref{fig:ghh} we report the RDFs computed among HDL-like oxygens only at 1 bar, in the temperature range $205\lesssim T\lesssim 235$K. Similar distributions characterize higher pressures. Upon cooling the sample, we observe an increase in the intensity of the first peak accompanied with a deepening of the first minimum located at $\sim3.2$ \AA. Moving beyond the SRO, we observe a marked peak at $\sim3.7$ \AA, whose intensity increases on cooling and whose position is mostly independent on the temperature. On the other hand, the consequent minimum only slightly decreases on cooling, but its position shifts from $\sim3.8$ \AA at $T=235$K to $\sim3.9$ \AA at $T=205$K. This peak accounts for the presence of interstitial molecules that populate the space between the first and the second shell of neighbours, and becomes more pronounced upon cooling, suggesting that the interstitial configuration become favourable upon reduction of the thermal energy. Interestingly, the intensity of this interstitial peak is comparable with the intensity of the main second peak at the lowest temperatures. The main second peak also shows a marked temperature dependence, becoming more pronounced upon cooling. The g$_{hh}$(r) become indistinguishable at radial distances of $r>6$ \AA, confirming that the HDL-like environments play and active role in the appearance of water anomalies. \\
In order to investigate in deeper detail the role of HDL-like environments, we have focused our attention on the interstitial molecules. While an inspection of the HBN suggests that the interstitial molecules are bonded to the central molecules at all thermodynamic conditions~\footnote{It is worthy to mention that in LDA, the interstitial molecules are not bonded to the central one~\cite{finney_2002,martelli_hyperuniformity}, suggesting that the HBN of LDA and LDL are substantially different, as reported in Ref.~\cite{martelli_hyperuniformity}}, we can acquire interesting information by computing the ratio, as a function of the temperature, between the intensity of the maximum of the interstitial peak (max$_{\textit{int}}$) and the corresponding minimum (min$_{\textit{int}}$). In fig.~\ref{fig:maxmin} we report the ratio max$_{\textit{int}}$/min$_{\textit{int}}$ computed at the three pressures and in the temperature range $190\lesssim T\lesssim235$K.
It is possible to recognize a change in slope in max$_{\textit{int}}$/min$_{\textit{int}}$ for all pressures as a function of the temperature. This change in slope is emphasized with dashed lines that serve as a visual guide. Remarkably, the change in slope occurs for all pressures in correspondence with the T$_{WL}$, suggesting that the spatial organization of HDL-like molecules plays an active role in water's anomalies. It is worthy to emphasize here that our analysis are limited by the relatively small samples, and by the small number of HDL-like molecules located at the interstitial configuration. Further inspections with larger statistics may help in better quantifying and framing the -hitherto unknown- role of interstitial molecules in the anomalies of water.\\
In fig.~\ref{fig:rings_hdl} we report P(n), the
normalized probability of having an n-folded ring connecting HDL-like environments only via their HBN. Panel (a) shows the temperature dependent distributions at 1 bar, panel (b) at 400 bar and panel (c) at 1000 bar. At variance with the LDL-like network, the P(n) for the HDL-like network does not show any interesting feature in correspondence with the WL. At T$>$T$_{WL}$, the distributions show a maximum at n6 and are broad, indicating the existence of longer rings to accommodate the higher density. Upon cooling the samples, the corresponding P(n) show a decrease in the number of longer rings and a corresponding increase in n5, reflecting the decrease in the thermal energy and the progressive fragmentation of the large HDL-like cluster into smaller domains.
\section{Conclusions}\label{conclusions}
We have explored the structural properties and the HBN of equilibrated supercooled liquid water described by the TIP4P/2005 interaction potential. We have inspected the behaviour of the SRO and of the IRO by employing the score function, and we have characterized local environments based on their LSI value. \\
We have inspected the composition, clustering and HBN of LDL-like and of HDL-like environments and we have shown that, \emph{in correspondence with the WL}:\\
(i) the percentage of LDL-like molecules is in the narrow window of $\sim20-30\%$ at \emph{all} the inspected pressures, indicating that this window may represent a "critical" concentration at which water shows anomalous behaviours;\\
(ii) a large, dominating HDL-like cluster undergoes fragmentation caused by the increasing number of LDL-like clusters that percolate and connect each other, in a spinodal-like decomposition scenario. We speculate here that such kinetics could be modeled with a modified Cahn-Hilliard equation~\cite{cahn_hilliard}, describing a phase separation occurring in correspondence with T$_{WL}$. We are currently making progresses in this direction;\\
(iii) the topology of the HBN within the LDL-like molecules is characterized by an equal number of pentagonal and hexagonal rings, creating a state of maximal frustration;\\
(iv) the ratio max$_{\textit{int}}$/min$_{\textit{int}}$ shows, upon cooling, a rapid change in slope, suggesting that the interstitial arrangement in HDL-like environments plays a central role in the emergence of water's anomalies;\\
(v) the IRO -as measured by the score function- shows a minimum in correspondence with T$_{WL}$ that decrease upon increasing the pressure, reflecting the enhanced density and entropy fluctuations upon approaching the hypothesized LLCP. \\ 
Our results indicate that water's anomalies are the combined effect of a delicate balance between HDL-like and LDL-like environments, how they cluster, how they form networks and how they organize in space. We have unbundled these effects, shedding light on the complex physics of water's anomalies. Beyond implications for water, our results may help in understanding the anomalous behaviour of other materials.
\begin{acknowledgments}
This work was supported by the STFC Hartree Centre's Innovation Return on Research programme, funded by the
Department for Business, Energy and Industrial Strategy.
\end{acknowledgments} 
\linespread{0.1}
\bibliography{aipsamp}

\begin{thebibliography}{116}%
\makeatletter
\providecommand \@ifxundefined [1]{%
 \@ifx{#1\undefined}
}%
\providecommand \@ifnum [1]{%
 \ifnum #1\expandafter \@firstoftwo
 \else \expandafter \@secondoftwo
 \fi
}%
\providecommand \@ifx [1]{%
 \ifx #1\expandafter \@firstoftwo
 \else \expandafter \@secondoftwo
 \fi
}%
\providecommand \natexlab [1]{#1}%
\providecommand \enquote  [1]{``#1''}%
\providecommand \bibnamefont  [1]{#1}%
\providecommand \bibfnamefont [1]{#1}%
\providecommand \citenamefont [1]{#1}%
\providecommand \href@noop [0]{\@secondoftwo}%
\providecommand \href [0]{\begingroup \@sanitize@url \@href}%
\providecommand \@href[1]{\@@startlink{#1}\@@href}%
\providecommand \@@href[1]{\endgroup#1\@@endlink}%
\providecommand \@sanitize@url [0]{\catcode `\\12\catcode `\$12\catcode
  `\&12\catcode `\#12\catcode `\^12\catcode `\_12\catcode `\%12\relax}%
\providecommand \@@startlink[1]{}%
\providecommand \@@endlink[0]{}%
\providecommand \url  [0]{\begingroup\@sanitize@url \@url }%
\providecommand \@url [1]{\endgroup\@href {#1}{\urlprefix }}%
\providecommand \urlprefix  [0]{URL }%
\providecommand \Eprint [0]{\href }%
\providecommand \doibase [0]{http://dx.doi.org/}%
\providecommand \selectlanguage [0]{\@gobble}%
\providecommand \bibinfo  [0]{\@secondoftwo}%
\providecommand \bibfield  [0]{\@secondoftwo}%
\providecommand \translation [1]{[#1]}%
\providecommand \BibitemOpen [0]{}%
\providecommand \bibitemStop [0]{}%
\providecommand \bibitemNoStop [0]{.\EOS\space}%
\providecommand \EOS [0]{\spacefactor3000\relax}%
\providecommand \BibitemShut  [1]{\csname bibitem#1\endcsname}%
\let\auto@bib@innerbib\@empty
\bibitem [{\citenamefont {Bernal}\ and\ \citenamefont
  {Fowler}(1933)}]{bernal_fowler}%
  \BibitemOpen
  \bibfield  {author} {\bibinfo {author} {\bibfnamefont {J.~D.}\ \bibnamefont
  {Bernal}}\ and\ \bibinfo {author} {\bibfnamefont {R.~H.}\ \bibnamefont
  {Fowler}},\ }\href@noop {} {\bibfield  {journal} {\bibinfo  {journal} {J.
  Chem. Phys.}\ }\textbf {\bibinfo {volume} {1}},\ \bibinfo {pages} {515}
  (\bibinfo {year} {1933})}\BibitemShut {NoStop}%
\bibitem [{\citenamefont {Pauling}(1960)}]{pauling}%
  \BibitemOpen
  \bibfield  {author} {\bibinfo {author} {\bibfnamefont {L.}~\bibnamefont
  {Pauling}},\ }\href@noop {} {\emph {\bibinfo {title} {The Nature of the
  Chemical Bond, and the Structure of Molecules and Crystals}}},\ \bibinfo
  {edition} {3rd}\ ed.\ (\bibinfo  {publisher} {Ithaca, NY: Cornell University
  Press},\ \bibinfo {year} {1960})\BibitemShut {NoStop}%
\bibitem [{\citenamefont {People}(1951)}]{people}%
  \BibitemOpen
  \bibfield  {author} {\bibinfo {author} {\bibfnamefont {J.~A.}\ \bibnamefont
  {People}},\ }\href@noop {} {\bibfield  {journal} {\bibinfo  {journal} {Proc.
  Royal Soc. A}\ }\textbf {\bibinfo {volume} {205}},\ \bibinfo {pages} {163}
  (\bibinfo {year} {1951})}\BibitemShut {NoStop}%
\bibitem [{\citenamefont {Angell}\ and\ \citenamefont
  {Angell}(1976)}]{speedy_angell}%
  \BibitemOpen
  \bibfield  {author} {\bibinfo {author} {\bibfnamefont {R.~J.}\ \bibnamefont
  {Angell}}\ and\ \bibinfo {author} {\bibfnamefont {C.~A.}\ \bibnamefont
  {Angell}},\ }\href@noop {} {\bibfield  {journal} {\bibinfo  {journal} {J.
  Chem. Phys/}\ }\textbf {\bibinfo {volume} {65}},\ \bibinfo {pages} {851}
  (\bibinfo {year} {1976})}\BibitemShut {NoStop}%
\bibitem [{\citenamefont {Poole}\ \emph {et~al.}(1992)\citenamefont {Poole},
  \citenamefont {Sciortino}, \citenamefont {Essmann},\ and\ \citenamefont
  {Stanley}}]{poole_nature}%
  \BibitemOpen
  \bibfield  {author} {\bibinfo {author} {\bibfnamefont {P.~H.}\ \bibnamefont
  {Poole}}, \bibinfo {author} {\bibfnamefont {F.}~\bibnamefont {Sciortino}},
  \bibinfo {author} {\bibfnamefont {U.}~\bibnamefont {Essmann}}, \ and\
  \bibinfo {author} {\bibfnamefont {H.~E.}\ \bibnamefont {Stanley}},\
  }\href@noop {} {\bibfield  {journal} {\bibinfo  {journal} {Nature}\ }\textbf
  {\bibinfo {volume} {360}},\ \bibinfo {pages} {324} (\bibinfo {year}
  {1992})}\BibitemShut {NoStop}%
\bibitem [{\citenamefont {Poole}\ \emph
  {et~al.}(1993{\natexlab{a}})\citenamefont {Poole}, \citenamefont {Sciortino},
  \citenamefont {Essmann},\ and\ \citenamefont {Stanley}}]{poole_pre_1}%
  \BibitemOpen
  \bibfield  {author} {\bibinfo {author} {\bibfnamefont {P.~H.}\ \bibnamefont
  {Poole}}, \bibinfo {author} {\bibfnamefont {F.}~\bibnamefont {Sciortino}},
  \bibinfo {author} {\bibfnamefont {U.}~\bibnamefont {Essmann}}, \ and\
  \bibinfo {author} {\bibfnamefont {H.~E.}\ \bibnamefont {Stanley}},\
  }\href@noop {} {\bibfield  {journal} {\bibinfo  {journal} {Phys. Rev. E}\
  }\textbf {\bibinfo {volume} {48}},\ \bibinfo {pages} {3799} (\bibinfo {year}
  {1993}{\natexlab{a}})}\BibitemShut {NoStop}%
\bibitem [{\citenamefont {Poole}\ \emph
  {et~al.}(1993{\natexlab{b}})\citenamefont {Poole}, \citenamefont {Sciortino},
  \citenamefont {Essmann},\ and\ \citenamefont {Stanley}}]{poole_pre_2}%
  \BibitemOpen
  \bibfield  {author} {\bibinfo {author} {\bibfnamefont {P.~H.}\ \bibnamefont
  {Poole}}, \bibinfo {author} {\bibfnamefont {F.}~\bibnamefont {Sciortino}},
  \bibinfo {author} {\bibfnamefont {U.}~\bibnamefont {Essmann}}, \ and\
  \bibinfo {author} {\bibfnamefont {H.~E.}\ \bibnamefont {Stanley}},\
  }\href@noop {} {\bibfield  {journal} {\bibinfo  {journal} {Phys. Rev. E}\
  }\textbf {\bibinfo {volume} {48}},\ \bibinfo {pages} {4605} (\bibinfo {year}
  {1993}{\natexlab{b}})}\BibitemShut {NoStop}%
\bibitem [{\citenamefont {Palmer}\ \emph {et~al.}(2014)\citenamefont {Palmer},
  \citenamefont {Martelli}, \citenamefont {Liu}, \citenamefont {Car},
  \citenamefont {Panagiotopoulos},\ and\ \citenamefont
  {Debenedetti}}]{mio_nature}%
  \BibitemOpen
  \bibfield  {author} {\bibinfo {author} {\bibfnamefont {J.~C.}\ \bibnamefont
  {Palmer}}, \bibinfo {author} {\bibfnamefont {F.}~\bibnamefont {Martelli}},
  \bibinfo {author} {\bibfnamefont {Y.}~\bibnamefont {Liu}}, \bibinfo {author}
  {\bibfnamefont {R.}~\bibnamefont {Car}}, \bibinfo {author} {\bibfnamefont
  {A.~Z.}\ \bibnamefont {Panagiotopoulos}}, \ and\ \bibinfo {author}
  {\bibfnamefont {P.~G.}\ \bibnamefont {Debenedetti}},\ }\href@noop {}
  {\bibfield  {journal} {\bibinfo  {journal} {Nature}\ }\textbf {\bibinfo
  {volume} {510}},\ \bibinfo {pages} {385} (\bibinfo {year}
  {2014})}\BibitemShut {NoStop}%
\bibitem [{\citenamefont {Xu}\ \emph {et~al.}(2005)\citenamefont {Xu},
  \citenamefont {Kumar}, \citenamefont {Buldyrev}, \citenamefont {Chen},
  \citenamefont {Poole}, \citenamefont {Sciortino},\ and\ \citenamefont
  {Stanley}}]{xu_pnas}%
  \BibitemOpen
  \bibfield  {author} {\bibinfo {author} {\bibfnamefont {L.}~\bibnamefont
  {Xu}}, \bibinfo {author} {\bibfnamefont {P.}~\bibnamefont {Kumar}}, \bibinfo
  {author} {\bibfnamefont {S.~V.}\ \bibnamefont {Buldyrev}}, \bibinfo {author}
  {\bibfnamefont {S.-H.}\ \bibnamefont {Chen}}, \bibinfo {author}
  {\bibfnamefont {P.~H.}\ \bibnamefont {Poole}}, \bibinfo {author}
  {\bibfnamefont {F.}~\bibnamefont {Sciortino}}, \ and\ \bibinfo {author}
  {\bibfnamefont {H.~E.}\ \bibnamefont {Stanley}},\ }\href@noop {} {\bibfield
  {journal} {\bibinfo  {journal} {Proc. Natl. Acad. Sci. USA}\ }\textbf
  {\bibinfo {volume} {102}},\ \bibinfo {pages} {16558} (\bibinfo {year}
  {2005})}\BibitemShut {NoStop}%
\bibitem [{\citenamefont {Abascal}\ and\ \citenamefont
  {Vega}(2010)}]{abascal_2010}%
  \BibitemOpen
  \bibfield  {author} {\bibinfo {author} {\bibfnamefont {J.~L.~F.}\
  \bibnamefont {Abascal}}\ and\ \bibinfo {author} {\bibfnamefont
  {C.}~\bibnamefont {Vega}},\ }\href@noop {} {\bibfield  {journal} {\bibinfo
  {journal} {J. Chem. Phys.}\ }\textbf {\bibinfo {volume} {133}},\ \bibinfo
  {pages} {234502} (\bibinfo {year} {2010})}\BibitemShut {NoStop}%
\bibitem [{\citenamefont {Liu}\ \emph {et~al.}(2010)\citenamefont {Liu},
  \citenamefont {Palmer}, \citenamefont {Panagiotopoulos},\ and\ \citenamefont
  {Debenedetti}}]{liu_2010}%
  \BibitemOpen
  \bibfield  {author} {\bibinfo {author} {\bibfnamefont {Y.}~\bibnamefont
  {Liu}}, \bibinfo {author} {\bibfnamefont {J.~C.}\ \bibnamefont {Palmer}},
  \bibinfo {author} {\bibfnamefont {A.~Z.}\ \bibnamefont {Panagiotopoulos}}, \
  and\ \bibinfo {author} {\bibfnamefont {P.~G.}\ \bibnamefont {Debenedetti}},\
  }\href@noop {} {\bibfield  {journal} {\bibinfo  {journal} {J. Chem. Phys.}\
  }\textbf {\bibinfo {volume} {137}},\ \bibinfo {pages} {214505} (\bibinfo
  {year} {2010})}\BibitemShut {NoStop}%
\bibitem [{\citenamefont {Abascal}\ and\ \citenamefont
  {Vega}(2011)}]{abascal_2011}%
  \BibitemOpen
  \bibfield  {author} {\bibinfo {author} {\bibfnamefont {J.}~\bibnamefont
  {Abascal}}\ and\ \bibinfo {author} {\bibfnamefont {C.}~\bibnamefont {Vega}},\
  }\href@noop {} {\bibfield  {journal} {\bibinfo  {journal} {J. Chem. Phys.}\
  }\textbf {\bibinfo {volume} {134}},\ \bibinfo {pages} {186101} (\bibinfo
  {year} {2011})}\BibitemShut {NoStop}%
\bibitem [{\citenamefont {Corradini}\ and\ \citenamefont
  {Gallo}(2011)}]{corradini_2011}%
  \BibitemOpen
  \bibfield  {author} {\bibinfo {author} {\bibfnamefont {D.}~\bibnamefont
  {Corradini}}\ and\ \bibinfo {author} {\bibfnamefont {P.}~\bibnamefont
  {Gallo}},\ }\href@noop {} {\bibfield  {journal} {\bibinfo  {journal} {J.
  Phys. Chem. B}\ }\textbf {\bibinfo {volume} {115}},\ \bibinfo {pages} {1461}
  (\bibinfo {year} {2011})}\BibitemShut {NoStop}%
\bibitem [{\citenamefont {Li}, \citenamefont {Li},\ and\ \citenamefont
  {Wang}(2013)}]{li_2013}%
  \BibitemOpen
  \bibfield  {author} {\bibinfo {author} {\bibfnamefont {Y.}~\bibnamefont
  {Li}}, \bibinfo {author} {\bibfnamefont {J.}~\bibnamefont {Li}}, \ and\
  \bibinfo {author} {\bibfnamefont {F.}~\bibnamefont {Wang}},\ }\href@noop {}
  {\bibfield  {journal} {\bibinfo  {journal} {Proc. Natl. Acad. Sci. USA}\
  }\textbf {\bibinfo {volume} {110}},\ \bibinfo {pages} {12209} (\bibinfo
  {year} {2013})}\BibitemShut {NoStop}%
\bibitem [{\citenamefont {Palmer}, \citenamefont {Car},\ and\ \citenamefont
  {Debenedetti}(2013)}]{palmer_2013}%
  \BibitemOpen
  \bibfield  {author} {\bibinfo {author} {\bibfnamefont {J.~C.}\ \bibnamefont
  {Palmer}}, \bibinfo {author} {\bibfnamefont {R.}~\bibnamefont {Car}}, \ and\
  \bibinfo {author} {\bibfnamefont {P.~G.}\ \bibnamefont {Debenedetti}},\
  }\href@noop {} {\bibfield  {journal} {\bibinfo  {journal} {Faraday Discuss.}\
  }\textbf {\bibinfo {volume} {167}},\ \bibinfo {pages} {77} (\bibinfo {year}
  {2013})}\BibitemShut {NoStop}%
\bibitem [{\citenamefont {Yagasaki}, \citenamefont {Matsumoto},\ and\
  \citenamefont {Tanaka}(2014)}]{yagasaki_2014}%
  \BibitemOpen
  \bibfield  {author} {\bibinfo {author} {\bibfnamefont {T.}~\bibnamefont
  {Yagasaki}}, \bibinfo {author} {\bibfnamefont {M.}~\bibnamefont {Matsumoto}},
  \ and\ \bibinfo {author} {\bibfnamefont {H.}~\bibnamefont {Tanaka}},\
  }\href@noop {} {\bibfield  {journal} {\bibinfo  {journal} {Phys. Rev. E}\
  }\textbf {\bibinfo {volume} {89}},\ \bibinfo {pages} {020301(R)} (\bibinfo
  {year} {2014})}\BibitemShut {NoStop}%
\bibitem [{\citenamefont {Holten}\ \emph {et~al.}(2014)\citenamefont {Holten},
  \citenamefont {Palmer}, \citenamefont {Poole}, \citenamefont {Debenedetti},\
  and\ \citenamefont {Anisimov}}]{holten_2014}%
  \BibitemOpen
  \bibfield  {author} {\bibinfo {author} {\bibfnamefont {V.}~\bibnamefont
  {Holten}}, \bibinfo {author} {\bibfnamefont {J.~C.}\ \bibnamefont {Palmer}},
  \bibinfo {author} {\bibfnamefont {P.~H.}\ \bibnamefont {Poole}}, \bibinfo
  {author} {\bibfnamefont {P.~G.}\ \bibnamefont {Debenedetti}}, \ and\ \bibinfo
  {author} {\bibfnamefont {M.~A.}\ \bibnamefont {Anisimov}},\ }\href@noop {}
  {\bibfield  {journal} {\bibinfo  {journal} {J. Chem. Phys.}\ }\textbf
  {\bibinfo {volume} {140}},\ \bibinfo {pages} {104502} (\bibinfo {year}
  {2014})}\BibitemShut {NoStop}%
\bibitem [{\citenamefont {Smallenburg}, \citenamefont {Filion},\ and\
  \citenamefont {Sciortino}(2014)}]{smallenburg_2014}%
  \BibitemOpen
  \bibfield  {author} {\bibinfo {author} {\bibfnamefont {F.}~\bibnamefont
  {Smallenburg}}, \bibinfo {author} {\bibfnamefont {L.}~\bibnamefont {Filion}},
  \ and\ \bibinfo {author} {\bibfnamefont {F.}~\bibnamefont {Sciortino}},\
  }\href@noop {} {\bibfield  {journal} {\bibinfo  {journal} {Nat. Phys.}\
  }\textbf {\bibinfo {volume} {10}},\ \bibinfo {pages} {653} (\bibinfo {year}
  {2014})}\BibitemShut {NoStop}%
\bibitem [{\citenamefont {Smallenburg}\ and\ \citenamefont
  {Sciortino}(2015)}]{smallenburg_2015}%
  \BibitemOpen
  \bibfield  {author} {\bibinfo {author} {\bibfnamefont {F.}~\bibnamefont
  {Smallenburg}}\ and\ \bibinfo {author} {\bibfnamefont {F.}~\bibnamefont
  {Sciortino}},\ }\href@noop {} {\bibfield  {journal} {\bibinfo  {journal}
  {Phys. Rev. Lett.}\ }\textbf {\bibinfo {volume} {115}},\ \bibinfo {pages}
  {015701} (\bibinfo {year} {2015})}\BibitemShut {NoStop}%
\bibitem [{\citenamefont {Ni}\ and\ \citenamefont {Skinner}(2016)}]{ni_2016}%
  \BibitemOpen
  \bibfield  {author} {\bibinfo {author} {\bibfnamefont {Y.}~\bibnamefont
  {Ni}}\ and\ \bibinfo {author} {\bibfnamefont {J.~L.}\ \bibnamefont
  {Skinner}},\ }\href@noop {} {\bibfield  {journal} {\bibinfo  {journal} {J.
  Chem. Phys.}\ }\textbf {\bibinfo {volume} {144}},\ \bibinfo {pages} {214501}
  (\bibinfo {year} {2016})}\BibitemShut {NoStop}%
\bibitem [{\citenamefont {Pathak}\ \emph {et~al.}(2016)\citenamefont {Pathak},
  \citenamefont {Palmer}, \citenamefont {Schlesinger}, \citenamefont
  {Wikfeldt}, \citenamefont {Sellberg}, \citenamefont {Pettersson},\ and\
  \citenamefont {Nilsson}}]{pathak_2016}%
  \BibitemOpen
  \bibfield  {author} {\bibinfo {author} {\bibfnamefont {H.}~\bibnamefont
  {Pathak}}, \bibinfo {author} {\bibfnamefont {J.~C.}\ \bibnamefont {Palmer}},
  \bibinfo {author} {\bibfnamefont {D.}~\bibnamefont {Schlesinger}}, \bibinfo
  {author} {\bibfnamefont {J.~T.}\ \bibnamefont {Wikfeldt}}, \bibinfo {author}
  {\bibfnamefont {J.~A.}\ \bibnamefont {Sellberg}}, \bibinfo {author}
  {\bibfnamefont {L.~G.}\ \bibnamefont {Pettersson}}, \ and\ \bibinfo {author}
  {\bibfnamefont {A.}~\bibnamefont {Nilsson}},\ }\href@noop {} {\bibfield
  {journal} {\bibinfo  {journal} {J. Chem. Phys.}\ }\textbf {\bibinfo {volume}
  {145}},\ \bibinfo {pages} {134507} (\bibinfo {year} {2016})}\BibitemShut
  {NoStop}%
\bibitem [{\citenamefont {Biddle}\ \emph {et~al.}(2017)\citenamefont {Biddle},
  \citenamefont {Singh}, \citenamefont {Sparano}, \citenamefont {Ricci},
  \citenamefont {González}, \citenamefont {Valeriani}, \citenamefont
  {Abascal}, \citenamefont {Debenedetti}, \citenamefont {Anisimov},\ and\
  \citenamefont {Caupin}}]{biddle_2017}%
  \BibitemOpen
  \bibfield  {author} {\bibinfo {author} {\bibfnamefont {J.~W.}\ \bibnamefont
  {Biddle}}, \bibinfo {author} {\bibfnamefont {R.~S.}\ \bibnamefont {Singh}},
  \bibinfo {author} {\bibfnamefont {E.~M.}\ \bibnamefont {Sparano}}, \bibinfo
  {author} {\bibfnamefont {F.}~\bibnamefont {Ricci}}, \bibinfo {author}
  {\bibfnamefont {M.~A.}\ \bibnamefont {González}}, \bibinfo {author}
  {\bibfnamefont {C.}~\bibnamefont {Valeriani}}, \bibinfo {author}
  {\bibfnamefont {J.~L.~F.}\ \bibnamefont {Abascal}}, \bibinfo {author}
  {\bibfnamefont {P.~G.}\ \bibnamefont {Debenedetti}}, \bibinfo {author}
  {\bibfnamefont {M.~A.}\ \bibnamefont {Anisimov}}, \ and\ \bibinfo {author}
  {\bibfnamefont {F.}~\bibnamefont {Caupin}},\ }\href@noop {} {\bibfield
  {journal} {\bibinfo  {journal} {J. Chem. Phys.}\ }\textbf {\bibinfo {volume}
  {146}},\ \bibinfo {pages} {034502} (\bibinfo {year} {2017})}\BibitemShut
  {NoStop}%
\bibitem [{\citenamefont {Palmer}\ \emph {et~al.}(2018)\citenamefont {Palmer},
  \citenamefont {Haji-Akbari}, \citenamefont {Singh}, \citenamefont {Martelli},
  \citenamefont {Car}, \citenamefont {Panagiotopoulos},\ and\ \citenamefont
  {Debenedetti}}]{palmer_2018}%
  \BibitemOpen
  \bibfield  {author} {\bibinfo {author} {\bibfnamefont {J.~C.}\ \bibnamefont
  {Palmer}}, \bibinfo {author} {\bibfnamefont {A.}~\bibnamefont {Haji-Akbari}},
  \bibinfo {author} {\bibfnamefont {R.~S.}\ \bibnamefont {Singh}}, \bibinfo
  {author} {\bibfnamefont {F.}~\bibnamefont {Martelli}}, \bibinfo {author}
  {\bibfnamefont {R.}~\bibnamefont {Car}}, \bibinfo {author} {\bibfnamefont
  {A.~Z.}\ \bibnamefont {Panagiotopoulos}}, \ and\ \bibinfo {author}
  {\bibfnamefont {P.~G.}\ \bibnamefont {Debenedetti}},\ }\href@noop {}
  {\bibfield  {journal} {\bibinfo  {journal} {J. Phys. Chem. B}\ }\textbf
  {\bibinfo {volume} {148}},\ \bibinfo {pages} {137101} (\bibinfo {year}
  {2018})}\BibitemShut {NoStop}%
\bibitem [{\citenamefont {Mishima}\ and\ \citenamefont
  {Suzuki}(2002)}]{mishima_2002}%
  \BibitemOpen
  \bibfield  {author} {\bibinfo {author} {\bibfnamefont {O.}~\bibnamefont
  {Mishima}}\ and\ \bibinfo {author} {\bibfnamefont {Y.}~\bibnamefont
  {Suzuki}},\ }\href@noop {} {\bibfield  {journal} {\bibinfo  {journal}
  {Nature}\ }\textbf {\bibinfo {volume} {419}},\ \bibinfo {pages} {599}
  (\bibinfo {year} {2002})}\BibitemShut {NoStop}%
\bibitem [{\citenamefont {Mishima}\ and\ \citenamefont
  {Stanley}(1998)}]{mishima_1998}%
  \BibitemOpen
  \bibfield  {author} {\bibinfo {author} {\bibfnamefont {O.}~\bibnamefont
  {Mishima}}\ and\ \bibinfo {author} {\bibfnamefont {H.~E.}\ \bibnamefont
  {Stanley}},\ }\href@noop {} {\bibfield  {journal} {\bibinfo  {journal}
  {Nature}\ }\textbf {\bibinfo {volume} {392}},\ \bibinfo {pages} {164}
  (\bibinfo {year} {1998})}\BibitemShut {NoStop}%
\bibitem [{\citenamefont {Mishima}(2000)}]{mishima_2000}%
  \BibitemOpen
  \bibfield  {author} {\bibinfo {author} {\bibfnamefont {O.}~\bibnamefont
  {Mishima}},\ }\href@noop {} {\bibfield  {journal} {\bibinfo  {journal} {Phys.
  Rev. Lett.}\ }\textbf {\bibinfo {volume} {85}},\ \bibinfo {pages} {334}
  (\bibinfo {year} {2000})}\BibitemShut {NoStop}%
\bibitem [{\citenamefont {Fuentevilla}\ and\ \citenamefont
  {Anisimov}(2006)}]{fuentevilla_2006}%
  \BibitemOpen
  \bibfield  {author} {\bibinfo {author} {\bibfnamefont {D.}~\bibnamefont
  {Fuentevilla}}\ and\ \bibinfo {author} {\bibfnamefont {M.}~\bibnamefont
  {Anisimov}},\ }\href@noop {} {\bibfield  {journal} {\bibinfo  {journal}
  {Phys. Rev. Lett.}\ }\textbf {\bibinfo {volume} {97}},\ \bibinfo {pages}
  {195702} (\bibinfo {year} {2006})}\BibitemShut {NoStop}%
\bibitem [{\citenamefont {Kanno}\ and\ \citenamefont
  {Miyata}(2006)}]{kanno_2006}%
  \BibitemOpen
  \bibfield  {author} {\bibinfo {author} {\bibfnamefont {H.}~\bibnamefont
  {Kanno}}\ and\ \bibinfo {author} {\bibfnamefont {K.}~\bibnamefont {Miyata}},\
  }\href@noop {} {\bibfield  {journal} {\bibinfo  {journal} {Chem. Phys.
  Lett.}\ }\textbf {\bibinfo {volume} {442}},\ \bibinfo {pages} {507} (\bibinfo
  {year} {2006})}\BibitemShut {NoStop}%
\bibitem [{\citenamefont {Mallamace}\ \emph {et~al.}(2008)\citenamefont
  {Mallamace}, \citenamefont {Corsaro}, \citenamefont {Broccio}, \citenamefont
  {Branca}, \citenamefont {Gonz\'ales-Segredo}, \citenamefont {Spooren},
  \citenamefont {Chen},\ and\ \citenamefont {Stanley}}]{mallamace_2008}%
  \BibitemOpen
  \bibfield  {author} {\bibinfo {author} {\bibfnamefont {F.}~\bibnamefont
  {Mallamace}}, \bibinfo {author} {\bibfnamefont {C.}~\bibnamefont {Corsaro}},
  \bibinfo {author} {\bibfnamefont {M.}~\bibnamefont {Broccio}}, \bibinfo
  {author} {\bibfnamefont {C.}~\bibnamefont {Branca}}, \bibinfo {author}
  {\bibfnamefont {N.}~\bibnamefont {Gonz\'ales-Segredo}}, \bibinfo {author}
  {\bibfnamefont {J.}~\bibnamefont {Spooren}}, \bibinfo {author} {\bibfnamefont
  {S.-H.}\ \bibnamefont {Chen}}, \ and\ \bibinfo {author} {\bibfnamefont
  {H.~E.}\ \bibnamefont {Stanley}},\ }\href@noop {} {\bibfield  {journal}
  {\bibinfo  {journal} {Proc. Natl. Acad. Sci. USA}\ }\textbf {\bibinfo
  {volume} {105}},\ \bibinfo {pages} {12725} (\bibinfo {year}
  {2008})}\BibitemShut {NoStop}%
\bibitem [{\citenamefont {Mishima}(2010)}]{mishima_2010}%
  \BibitemOpen
  \bibfield  {author} {\bibinfo {author} {\bibfnamefont {O.}~\bibnamefont
  {Mishima}},\ }\href@noop {} {\bibfield  {journal} {\bibinfo  {journal} {J.
  Chem. Phys.}\ }\textbf {\bibinfo {volume} {133}},\ \bibinfo {pages} {144503}
  (\bibinfo {year} {2010})}\BibitemShut {NoStop}%
\bibitem [{\citenamefont {Bertrand}\ and\ \citenamefont
  {Anisimov}(2011)}]{bertrand_2011}%
  \BibitemOpen
  \bibfield  {author} {\bibinfo {author} {\bibfnamefont {C.}~\bibnamefont
  {Bertrand}}\ and\ \bibinfo {author} {\bibfnamefont {M.}~\bibnamefont
  {Anisimov}},\ }\href@noop {} {\bibfield  {journal} {\bibinfo  {journal} {J.
  Phys. Chem. B}\ }\textbf {\bibinfo {volume} {115}},\ \bibinfo {pages} {14099}
  (\bibinfo {year} {2011})}\BibitemShut {NoStop}%
\bibitem [{\citenamefont {Holten}\ \emph {et~al.}(2012)\citenamefont {Holten},
  \citenamefont {Bertrand}, \citenamefont {Anisimov},\ and\ \citenamefont
  {Sengers}}]{holten_2012_2}%
  \BibitemOpen
  \bibfield  {author} {\bibinfo {author} {\bibfnamefont {V.}~\bibnamefont
  {Holten}}, \bibinfo {author} {\bibfnamefont {C.}~\bibnamefont {Bertrand}},
  \bibinfo {author} {\bibfnamefont {M.}~\bibnamefont {Anisimov}}, \ and\
  \bibinfo {author} {\bibfnamefont {J.}~\bibnamefont {Sengers}},\ }\href@noop
  {} {\bibfield  {journal} {\bibinfo  {journal} {J. Chem. Phys.}\ }\textbf
  {\bibinfo {volume} {136}},\ \bibinfo {pages} {094507} (\bibinfo {year}
  {2012})}\BibitemShut {NoStop}%
\bibitem [{\citenamefont {Sellberg}\ \emph
  {et~al.}(2014{\natexlab{a}})\citenamefont {Sellberg}, \citenamefont {Kaya},
  \citenamefont {Segtnan}, \citenamefont {Chen}, \citenamefont {Tyliszczak},
  \citenamefont {Ogasawara}, \citenamefont {Nordlund}, \citenamefont
  {Pettersson},\ and\ \citenamefont {Nilsson}}]{sellberg_2014}%
  \BibitemOpen
  \bibfield  {author} {\bibinfo {author} {\bibfnamefont {J.~A.}\ \bibnamefont
  {Sellberg}}, \bibinfo {author} {\bibfnamefont {S.}~\bibnamefont {Kaya}},
  \bibinfo {author} {\bibfnamefont {V.~H.}\ \bibnamefont {Segtnan}}, \bibinfo
  {author} {\bibfnamefont {C.}~\bibnamefont {Chen}}, \bibinfo {author}
  {\bibfnamefont {T.}~\bibnamefont {Tyliszczak}}, \bibinfo {author}
  {\bibfnamefont {H.}~\bibnamefont {Ogasawara}}, \bibinfo {author}
  {\bibfnamefont {D.}~\bibnamefont {Nordlund}}, \bibinfo {author}
  {\bibfnamefont {L.~G.~M.}\ \bibnamefont {Pettersson}}, \ and\ \bibinfo
  {author} {\bibfnamefont {A.}~\bibnamefont {Nilsson}},\ }\href@noop {}
  {\bibfield  {journal} {\bibinfo  {journal} {J. Chem. Phys.}\ }\textbf
  {\bibinfo {volume} {141}},\ \bibinfo {pages} {034507} (\bibinfo {year}
  {2014}{\natexlab{a}})}\BibitemShut {NoStop}%
\bibitem [{\citenamefont {Sellberg}\ \emph
  {et~al.}(2014{\natexlab{b}})\citenamefont {Sellberg}, \citenamefont {Huang},
  \citenamefont {McQueen}, \citenamefont {Loh}, \citenamefont {Laksmono},
  \citenamefont {Sclesinger}, \citenamefont {Sierra}, \citenamefont {Nordlund},
  \citenamefont {Hampton}, \citenamefont {Starodub}, \citenamefont {DePonte},
  \citenamefont {Beye}, \citenamefont {Chen}, \citenamefont {Martin},
  \citenamefont {Barty}, \citenamefont {Wikfeldt}, \citenamefont {Weiss},
  \citenamefont {Caronna}, \citenamefont {Feldkamp}, \citenamefont {Skinner},
  \citenamefont {Seibert}, \citenamefont {Messerschmidt}, \citenamefont
  {Williams}, \citenamefont {Boutet}, \citenamefont {Pettersson}, \citenamefont
  {Bogan},\ and\ \citenamefont {Nilsson}}]{sellberg_2014_nature}%
  \BibitemOpen
  \bibfield  {author} {\bibinfo {author} {\bibfnamefont {J.~A.}\ \bibnamefont
  {Sellberg}}, \bibinfo {author} {\bibfnamefont {C.}~\bibnamefont {Huang}},
  \bibinfo {author} {\bibfnamefont {T.~A.}\ \bibnamefont {McQueen}}, \bibinfo
  {author} {\bibfnamefont {N.~D.}\ \bibnamefont {Loh}}, \bibinfo {author}
  {\bibfnamefont {H.}~\bibnamefont {Laksmono}}, \bibinfo {author}
  {\bibfnamefont {D.}~\bibnamefont {Sclesinger}}, \bibinfo {author}
  {\bibfnamefont {R.~G.}\ \bibnamefont {Sierra}}, \bibinfo {author}
  {\bibfnamefont {D.}~\bibnamefont {Nordlund}}, \bibinfo {author}
  {\bibfnamefont {C.~Y.}\ \bibnamefont {Hampton}}, \bibinfo {author}
  {\bibfnamefont {D.}~\bibnamefont {Starodub}}, \bibinfo {author}
  {\bibfnamefont {D.~P.}\ \bibnamefont {DePonte}}, \bibinfo {author}
  {\bibfnamefont {M.}~\bibnamefont {Beye}}, \bibinfo {author} {\bibfnamefont
  {C.}~\bibnamefont {Chen}}, \bibinfo {author} {\bibfnamefont {A.~V.}\
  \bibnamefont {Martin}}, \bibinfo {author} {\bibfnamefont {A.}~\bibnamefont
  {Barty}}, \bibinfo {author} {\bibfnamefont {K.~T.}\ \bibnamefont {Wikfeldt}},
  \bibinfo {author} {\bibfnamefont {T.~M.}\ \bibnamefont {Weiss}}, \bibinfo
  {author} {\bibfnamefont {C.}~\bibnamefont {Caronna}}, \bibinfo {author}
  {\bibfnamefont {J.}~\bibnamefont {Feldkamp}}, \bibinfo {author}
  {\bibfnamefont {L.~B.}\ \bibnamefont {Skinner}}, \bibinfo {author}
  {\bibfnamefont {M.~M.}\ \bibnamefont {Seibert}}, \bibinfo {author}
  {\bibfnamefont {M.}~\bibnamefont {Messerschmidt}}, \bibinfo {author}
  {\bibfnamefont {G.~J.}\ \bibnamefont {Williams}}, \bibinfo {author}
  {\bibfnamefont {S.}~\bibnamefont {Boutet}}, \bibinfo {author} {\bibfnamefont
  {L.~G.~M.}\ \bibnamefont {Pettersson}}, \bibinfo {author} {\bibfnamefont
  {M.~J.}\ \bibnamefont {Bogan}}, \ and\ \bibinfo {author} {\bibfnamefont
  {A.}~\bibnamefont {Nilsson}},\ }\href@noop {} {\bibfield  {journal} {\bibinfo
   {journal} {Nature}\ }\textbf {\bibinfo {volume} {510}},\ \bibinfo {pages}
  {381} (\bibinfo {year} {2014}{\natexlab{b}})}\BibitemShut {NoStop}%
\bibitem [{\citenamefont {Fanetti}\ \emph
  {et~al.}(2014{\natexlab{a}})\citenamefont {Fanetti}, \citenamefont {Pagliai},
  \citenamefont {Citroni}, \citenamefont {Lapini}, \citenamefont {Scandolo},
  \citenamefont {Righini},\ and\ \citenamefont {Bini}}]{fanetti_2014_2}%
  \BibitemOpen
  \bibfield  {author} {\bibinfo {author} {\bibfnamefont {S.}~\bibnamefont
  {Fanetti}}, \bibinfo {author} {\bibfnamefont {M.}~\bibnamefont {Pagliai}},
  \bibinfo {author} {\bibfnamefont {M.}~\bibnamefont {Citroni}}, \bibinfo
  {author} {\bibfnamefont {A.}~\bibnamefont {Lapini}}, \bibinfo {author}
  {\bibfnamefont {S.}~\bibnamefont {Scandolo}}, \bibinfo {author}
  {\bibfnamefont {R.}~\bibnamefont {Righini}}, \ and\ \bibinfo {author}
  {\bibfnamefont {R.}~\bibnamefont {Bini}},\ }\href@noop {} {\bibfield
  {journal} {\bibinfo  {journal} {J. Phys. Chem. Lett.}\ }\textbf {\bibinfo
  {volume} {5}},\ \bibinfo {pages} {3804} (\bibinfo {year}
  {2014}{\natexlab{a}})}\BibitemShut {NoStop}%
\bibitem [{\citenamefont {Kim}\ \emph {et~al.}(2017)\citenamefont {Kim},
  \citenamefont {Sp\"ah}, \citenamefont {Pathak}, \citenamefont {Perakis},
  \citenamefont {Mariedahl}, \citenamefont {Amann-Winkel}, \citenamefont
  {Sellberg}, \citenamefont {Lee}, \citenamefont {Kim}, \citenamefont {Park},
  \citenamefont {Nam}, \citenamefont {Katayama},\ and\ \citenamefont
  {Nilsson}}]{kim_2017}%
  \BibitemOpen
  \bibfield  {author} {\bibinfo {author} {\bibfnamefont {K.~H.}\ \bibnamefont
  {Kim}}, \bibinfo {author} {\bibfnamefont {A.}~\bibnamefont {Sp\"ah}},
  \bibinfo {author} {\bibfnamefont {H.}~\bibnamefont {Pathak}}, \bibinfo
  {author} {\bibfnamefont {F.}~\bibnamefont {Perakis}}, \bibinfo {author}
  {\bibfnamefont {D.}~\bibnamefont {Mariedahl}}, \bibinfo {author}
  {\bibfnamefont {K.}~\bibnamefont {Amann-Winkel}}, \bibinfo {author}
  {\bibfnamefont {J.~A.}\ \bibnamefont {Sellberg}}, \bibinfo {author}
  {\bibfnamefont {J.~H.}\ \bibnamefont {Lee}}, \bibinfo {author} {\bibfnamefont
  {S.}~\bibnamefont {Kim}}, \bibinfo {author} {\bibfnamefont {J.}~\bibnamefont
  {Park}}, \bibinfo {author} {\bibfnamefont {K.~H.}\ \bibnamefont {Nam}},
  \bibinfo {author} {\bibfnamefont {T.}~\bibnamefont {Katayama}}, \ and\
  \bibinfo {author} {\bibfnamefont {A.}~\bibnamefont {Nilsson}},\ }\href@noop
  {} {\bibfield  {journal} {\bibinfo  {journal} {Science}\ }\textbf {\bibinfo
  {volume} {358}},\ \bibinfo {pages} {1589} (\bibinfo {year}
  {2017})}\BibitemShut {NoStop}%
\bibitem [{\citenamefont {Mishima}, \citenamefont {Calvert},\ and\
  \citenamefont {Whalley}(1985)}]{mishima_1985}%
  \BibitemOpen
  \bibfield  {author} {\bibinfo {author} {\bibfnamefont {O.}~\bibnamefont
  {Mishima}}, \bibinfo {author} {\bibfnamefont {L.}~\bibnamefont {Calvert}}, \
  and\ \bibinfo {author} {\bibfnamefont {E.}~\bibnamefont {Whalley}},\
  }\href@noop {} {\bibfield  {journal} {\bibinfo  {journal} {Nature}\ }\textbf
  {\bibinfo {volume} {324}},\ \bibinfo {pages} {76} (\bibinfo {year}
  {1985})}\BibitemShut {NoStop}%
\bibitem [{\citenamefont {Mishima}(1994)}]{mishima_1994}%
  \BibitemOpen
  \bibfield  {author} {\bibinfo {author} {\bibfnamefont {O.}~\bibnamefont
  {Mishima}},\ }\href@noop {} {\bibfield  {journal} {\bibinfo  {journal} {J.
  Chem. Phys.}\ }\textbf {\bibinfo {volume} {100}},\ \bibinfo {pages} {5910}
  (\bibinfo {year} {1994})}\BibitemShut {NoStop}%
\bibitem [{\citenamefont {Klotz}\ \emph {et~al.}(2005)\citenamefont {Klotz},
  \citenamefont {Str\"{a}sse}, \citenamefont {Nelmes}, \citenamefont {Loveday},
  \citenamefont {Hamel}, \citenamefont {Rousse}, \citenamefont {Canny},
  \citenamefont {Chervin},\ and\ \citenamefont {Saitta}}]{klotz)2005}%
  \BibitemOpen
  \bibfield  {author} {\bibinfo {author} {\bibfnamefont {S.}~\bibnamefont
  {Klotz}}, \bibinfo {author} {\bibfnamefont {T.}~\bibnamefont {Str\"{a}sse}},
  \bibinfo {author} {\bibfnamefont {R.}~\bibnamefont {Nelmes}}, \bibinfo
  {author} {\bibfnamefont {J.}~\bibnamefont {Loveday}}, \bibinfo {author}
  {\bibfnamefont {G.}~\bibnamefont {Hamel}}, \bibinfo {author} {\bibfnamefont
  {G.}~\bibnamefont {Rousse}}, \bibinfo {author} {\bibfnamefont
  {B.}~\bibnamefont {Canny}}, \bibinfo {author} {\bibfnamefont
  {J.}~\bibnamefont {Chervin}}, \ and\ \bibinfo {author} {\bibfnamefont
  {A.}~\bibnamefont {Saitta}},\ }\href@noop {} {\bibfield  {journal} {\bibinfo
  {journal} {Phys. Rev. Lett.}\ }\textbf {\bibinfo {volume} {94}},\ \bibinfo
  {pages} {025506} (\bibinfo {year} {2005})}\BibitemShut {NoStop}%
\bibitem [{\citenamefont {Winkel}, \citenamefont {Mayer},\ and\ \citenamefont
  {Loerting}(2011)}]{winkel_2011}%
  \BibitemOpen
  \bibfield  {author} {\bibinfo {author} {\bibfnamefont {K.}~\bibnamefont
  {Winkel}}, \bibinfo {author} {\bibfnamefont {E.}~\bibnamefont {Mayer}}, \
  and\ \bibinfo {author} {\bibfnamefont {T.}~\bibnamefont {Loerting}},\
  }\href@noop {} {\bibfield  {journal} {\bibinfo  {journal} {J. Phys. Chem. B}\
  }\textbf {\bibinfo {volume} {115}},\ \bibinfo {pages} {14141} (\bibinfo
  {year} {2011})}\BibitemShut {NoStop}%
\bibitem [{\citenamefont {Martelli}\ \emph {et~al.}(2017)\citenamefont
  {Martelli}, \citenamefont {Torquato}, \citenamefont {Giovambattista},\ and\
  \citenamefont {Car}}]{martelli_hyperuniformity}%
  \BibitemOpen
  \bibfield  {author} {\bibinfo {author} {\bibfnamefont {F.}~\bibnamefont
  {Martelli}}, \bibinfo {author} {\bibfnamefont {S.}~\bibnamefont {Torquato}},
  \bibinfo {author} {\bibfnamefont {N.}~\bibnamefont {Giovambattista}}, \ and\
  \bibinfo {author} {\bibfnamefont {R.}~\bibnamefont {Car}},\ }\href@noop {}
  {\bibfield  {journal} {\bibinfo  {journal} {Phys. Rev. Lett.}\ }\textbf
  {\bibinfo {volume} {119}},\ \bibinfo {pages} {136002} (\bibinfo {year}
  {2017})}\BibitemShut {NoStop}%
\bibitem [{\citenamefont {Martelli}\ \emph
  {et~al.}(2018{\natexlab{a}})\citenamefont {Martelli}, \citenamefont
  {Giovambattista}, \citenamefont {Torquato},\ and\ \citenamefont
  {Car}}]{martelli_searching}%
  \BibitemOpen
  \bibfield  {author} {\bibinfo {author} {\bibfnamefont {F.}~\bibnamefont
  {Martelli}}, \bibinfo {author} {\bibfnamefont {N.}~\bibnamefont
  {Giovambattista}}, \bibinfo {author} {\bibfnamefont {S.}~\bibnamefont
  {Torquato}}, \ and\ \bibinfo {author} {\bibfnamefont {R.}~\bibnamefont
  {Car}},\ }\href@noop {} {\bibfield  {journal} {\bibinfo  {journal} {Phys.
  Rev. Materials}\ }\textbf {\bibinfo {volume} {2}},\ \bibinfo {pages} {075601}
  (\bibinfo {year} {2018}{\natexlab{a}})}\BibitemShut {NoStop}%
\bibitem [{\citenamefont {Loerting}\ \emph {et~al.}(2001)\citenamefont
  {Loerting}, \citenamefont {Salzmann}, \citenamefont {Kohl}, \citenamefont
  {Mayer},\ and\ \citenamefont {Hallbrucker}}]{loerting_2001}%
  \BibitemOpen
  \bibfield  {author} {\bibinfo {author} {\bibfnamefont {T.}~\bibnamefont
  {Loerting}}, \bibinfo {author} {\bibfnamefont {C.}~\bibnamefont {Salzmann}},
  \bibinfo {author} {\bibfnamefont {I.}~\bibnamefont {Kohl}}, \bibinfo {author}
  {\bibfnamefont {E.}~\bibnamefont {Mayer}}, \ and\ \bibinfo {author}
  {\bibfnamefont {A.}~\bibnamefont {Hallbrucker}},\ }\href@noop {} {\bibfield
  {journal} {\bibinfo  {journal} {Phys. Chem. Chem. Phys.}\ }\textbf {\bibinfo
  {volume} {3}},\ \bibinfo {pages} {5355} (\bibinfo {year} {2001})}\BibitemShut
  {NoStop}%
\bibitem [{\citenamefont {Stillinger}\ and\ \citenamefont
  {Weber}(1984{\natexlab{a}})}]{stillinger_1984_1}%
  \BibitemOpen
  \bibfield  {author} {\bibinfo {author} {\bibfnamefont {F.~H.}\ \bibnamefont
  {Stillinger}}\ and\ \bibinfo {author} {\bibfnamefont {T.~A.}\ \bibnamefont
  {Weber}},\ }\href@noop {} {\bibfield  {journal} {\bibinfo  {journal}
  {Science}\ }\textbf {\bibinfo {volume} {225}},\ \bibinfo {pages} {983}
  (\bibinfo {year} {1984}{\natexlab{a}})}\BibitemShut {NoStop}%
\bibitem [{\citenamefont {Stillinger}\ and\ \citenamefont
  {Weber}(1984{\natexlab{b}})}]{stillinger_1984_2}%
  \BibitemOpen
  \bibfield  {author} {\bibinfo {author} {\bibfnamefont {F.~H.}\ \bibnamefont
  {Stillinger}}\ and\ \bibinfo {author} {\bibfnamefont {T.~A.}\ \bibnamefont
  {Weber}},\ }\href@noop {} {\bibfield  {journal} {\bibinfo  {journal} {J.
  Chem. Phys.}\ }\textbf {\bibinfo {volume} {80}},\ \bibinfo {pages} {4434}
  (\bibinfo {year} {1984}{\natexlab{b}})}\BibitemShut {NoStop}%
\bibitem [{\citenamefont {Santra}\ \emph {et~al.}(2015)\citenamefont {Santra},
  \citenamefont {Jr.}, \citenamefont {Martelli},\ and\ \citenamefont
  {Car}}]{santra_2015}%
  \BibitemOpen
  \bibfield  {author} {\bibinfo {author} {\bibfnamefont {B.}~\bibnamefont
  {Santra}}, \bibinfo {author} {\bibfnamefont {R.~A.~D.}\ \bibnamefont {Jr.}},
  \bibinfo {author} {\bibfnamefont {F.}~\bibnamefont {Martelli}}, \ and\
  \bibinfo {author} {\bibfnamefont {R.}~\bibnamefont {Car}},\ }\href@noop {}
  {\bibfield  {journal} {\bibinfo  {journal} {Mol. Phys.}\ }\textbf {\bibinfo
  {volume} {113}},\ \bibinfo {pages} {2829} (\bibinfo {year}
  {2015})}\BibitemShut {NoStop}%
\bibitem [{\citenamefont {Taschin}\ \emph {et~al.}(2013)\citenamefont
  {Taschin}, \citenamefont {Bartolini}, \citenamefont {Eramo}, \citenamefont
  {Righini},\ and\ \citenamefont {Torre}}]{torre_2013}%
  \BibitemOpen
  \bibfield  {author} {\bibinfo {author} {\bibfnamefont {A.}~\bibnamefont
  {Taschin}}, \bibinfo {author} {\bibfnamefont {P.}~\bibnamefont {Bartolini}},
  \bibinfo {author} {\bibfnamefont {R.}~\bibnamefont {Eramo}}, \bibinfo
  {author} {\bibfnamefont {R.}~\bibnamefont {Righini}}, \ and\ \bibinfo
  {author} {\bibfnamefont {R.}~\bibnamefont {Torre}},\ }\href@noop {}
  {\bibfield  {journal} {\bibinfo  {journal} {Nat. Comm.}\ }\textbf {\bibinfo
  {volume} {4}},\ \bibinfo {pages} {2301} (\bibinfo {year} {2013})}\BibitemShut
  {NoStop}%
\bibitem [{\citenamefont {Fanetti}\ \emph
  {et~al.}(2014{\natexlab{b}})\citenamefont {Fanetti}, \citenamefont {Lapini},
  \citenamefont {Pagliai}, \citenamefont {Citroni}, \citenamefont {Donato},
  \citenamefont {Scandolo}, \citenamefont {Righini},\ and\ \citenamefont
  {Bini}}]{fanetti_2014_1}%
  \BibitemOpen
  \bibfield  {author} {\bibinfo {author} {\bibfnamefont {S.}~\bibnamefont
  {Fanetti}}, \bibinfo {author} {\bibfnamefont {A.}~\bibnamefont {Lapini}},
  \bibinfo {author} {\bibfnamefont {M.}~\bibnamefont {Pagliai}}, \bibinfo
  {author} {\bibfnamefont {M.}~\bibnamefont {Citroni}}, \bibinfo {author}
  {\bibfnamefont {M.~D.}\ \bibnamefont {Donato}}, \bibinfo {author}
  {\bibfnamefont {S.}~\bibnamefont {Scandolo}}, \bibinfo {author}
  {\bibfnamefont {R.}~\bibnamefont {Righini}}, \ and\ \bibinfo {author}
  {\bibfnamefont {R.}~\bibnamefont {Bini}},\ }\href@noop {} {\bibfield
  {journal} {\bibinfo  {journal} {J. Phys. Chem. Lett.}\ }\textbf {\bibinfo
  {volume} {5}},\ \bibinfo {pages} {235} (\bibinfo {year}
  {2014}{\natexlab{b}})}\BibitemShut {NoStop}%
\bibitem [{\citenamefont {Tanaka}(2000{\natexlab{a}})}]{tanaka_2000}%
  \BibitemOpen
  \bibfield  {author} {\bibinfo {author} {\bibfnamefont {H.}~\bibnamefont
  {Tanaka}},\ }\href@noop {} {\bibfield  {journal} {\bibinfo  {journal} {J.
  Chem. Phys.}\ }\textbf {\bibinfo {volume} {112}},\ \bibinfo {pages} {799}
  (\bibinfo {year} {2000}{\natexlab{a}})}\BibitemShut {NoStop}%
\bibitem [{\citenamefont {Tanaka}(2000{\natexlab{b}})}]{tanaka_2000_2}%
  \BibitemOpen
  \bibfield  {author} {\bibinfo {author} {\bibfnamefont {H.}~\bibnamefont
  {Tanaka}},\ }\href@noop {} {\bibfield  {journal} {\bibinfo  {journal}
  {Europhys. Lett.}\ }\textbf {\bibinfo {volume} {50}},\ \bibinfo {pages} {340}
  (\bibinfo {year} {2000}{\natexlab{b}})}\BibitemShut {NoStop}%
\bibitem [{\citenamefont {Russo}\ and\ \citenamefont
  {Tanaka}(2014)}]{russo_2014}%
  \BibitemOpen
  \bibfield  {author} {\bibinfo {author} {\bibfnamefont {J.}~\bibnamefont
  {Russo}}\ and\ \bibinfo {author} {\bibfnamefont {H.}~\bibnamefont {Tanaka}},\
  }\href@noop {} {\bibfield  {journal} {\bibinfo  {journal} {Nat. Commun.}\
  }\textbf {\bibinfo {volume} {5}},\ \bibinfo {pages} {3556} (\bibinfo {year}
  {2014})}\BibitemShut {NoStop}%
\bibitem [{\citenamefont {Holten}\ and\ \citenamefont
  {Anisimov}(2012)}]{holten_2012}%
  \BibitemOpen
  \bibfield  {author} {\bibinfo {author} {\bibfnamefont {V.}~\bibnamefont
  {Holten}}\ and\ \bibinfo {author} {\bibfnamefont {M.~A.}\ \bibnamefont
  {Anisimov}},\ }\href@noop {} {\bibfield  {journal} {\bibinfo  {journal} {Sci.
  Rep.}\ }\textbf {\bibinfo {volume} {2}},\ \bibinfo {pages} {713} (\bibinfo
  {year} {2012})}\BibitemShut {NoStop}%
\bibitem [{\citenamefont {Nilsson}\ and\ \citenamefont
  {Pettersson}(2015)}]{nilsson_2015}%
  \BibitemOpen
  \bibfield  {author} {\bibinfo {author} {\bibfnamefont {A.}~\bibnamefont
  {Nilsson}}\ and\ \bibinfo {author} {\bibfnamefont {L.~G.~M.}\ \bibnamefont
  {Pettersson}},\ }\href@noop {} {\bibfield  {journal} {\bibinfo  {journal}
  {Nat. Commun.}\ }\textbf {\bibinfo {volume} {6}},\ \bibinfo {pages} {8998}
  (\bibinfo {year} {2015})}\BibitemShut {NoStop}%
\bibitem [{\citenamefont {Anisimov}(2018)}]{anisimov_2018}%
  \BibitemOpen
  \bibfield  {author} {\bibinfo {author} {\bibfnamefont {M.~A.}\ \bibnamefont
  {Anisimov}},\ }\href@noop {} {\bibfield  {journal} {\bibinfo  {journal}
  {Phys. Rev. X}\ }\textbf {\bibinfo {volume} {8}},\ \bibinfo {pages} {011004}
  (\bibinfo {year} {2018})}\BibitemShut {NoStop}%
\bibitem [{\citenamefont {Walrafen}(1964)}]{walrafen_1964}%
  \BibitemOpen
  \bibfield  {author} {\bibinfo {author} {\bibfnamefont {G.}~\bibnamefont
  {Walrafen}},\ }\href@noop {} {\bibfield  {journal} {\bibinfo  {journal} {J.
  Chem. Phys.}\ }\textbf {\bibinfo {volume} {40}},\ \bibinfo {pages} {3249}
  (\bibinfo {year} {1964})}\BibitemShut {NoStop}%
\bibitem [{\citenamefont {Walrafen}(1967)}]{walrafen_1967}%
  \BibitemOpen
  \bibfield  {author} {\bibinfo {author} {\bibfnamefont {G.}~\bibnamefont
  {Walrafen}},\ }\href@noop {} {\bibfield  {journal} {\bibinfo  {journal} {J.
  Chem. Phys.}\ }\textbf {\bibinfo {volume} {47}},\ \bibinfo {pages} {114}
  (\bibinfo {year} {1967})}\BibitemShut {NoStop}%
\bibitem [{\citenamefont {Walrafen}\ \emph {et~al.}(1986)\citenamefont
  {Walrafen}, \citenamefont {Fisher}, \citenamefont {Hokmabadi},\ and\
  \citenamefont {Yang}}]{walrafen_1986}%
  \BibitemOpen
  \bibfield  {author} {\bibinfo {author} {\bibfnamefont {G.}~\bibnamefont
  {Walrafen}}, \bibinfo {author} {\bibfnamefont {M.}~\bibnamefont {Fisher}},
  \bibinfo {author} {\bibfnamefont {M.}~\bibnamefont {Hokmabadi}}, \ and\
  \bibinfo {author} {\bibfnamefont {W.~H.}\ \bibnamefont {Yang}},\ }\href@noop
  {} {\bibfield  {journal} {\bibinfo  {journal} {J. Chem. Phys.}\ }\textbf
  {\bibinfo {volume} {85}},\ \bibinfo {pages} {6970} (\bibinfo {year}
  {1986})}\BibitemShut {NoStop}%
\bibitem [{\citenamefont {Woutersen}, \citenamefont {Emmerichs},\ and\
  \citenamefont {Bakker}(1997)}]{woutersen_1997}%
  \BibitemOpen
  \bibfield  {author} {\bibinfo {author} {\bibfnamefont {S.}~\bibnamefont
  {Woutersen}}, \bibinfo {author} {\bibfnamefont {U.}~\bibnamefont
  {Emmerichs}}, \ and\ \bibinfo {author} {\bibfnamefont {H.}~\bibnamefont
  {Bakker}},\ }\href@noop {} {\bibfield  {journal} {\bibinfo  {journal}
  {Scienc}\ }\textbf {\bibinfo {volume} {278}},\ \bibinfo {pages} {658}
  (\bibinfo {year} {1997})}\BibitemShut {NoStop}%
\bibitem [{\citenamefont {Huang}\ \emph {et~al.}(2009)\citenamefont {Huang},
  \citenamefont {Wikfeldt}, \citenamefont {Tokushima}, \citenamefont
  {Nordlund}, \citenamefont {Harada}, \citenamefont {Bergmann}, \citenamefont
  {Niebuhr}, \citenamefont {Weiss}, \citenamefont {Horikawa}, \citenamefont
  {Leetmaa}, \citenamefont {Ljungberg}, \citenamefont {Takahashi},
  \citenamefont {Lenz}, \citenamefont {Ojam\"{a}e}, \citenamefont {Lyubartsev},
  \citenamefont {Shin}, \citenamefont {Pettersson},\ and\ \citenamefont
  {Nilsson}}]{huang_2009}%
  \BibitemOpen
  \bibfield  {author} {\bibinfo {author} {\bibfnamefont {C.}~\bibnamefont
  {Huang}}, \bibinfo {author} {\bibfnamefont {K.~T.}\ \bibnamefont {Wikfeldt}},
  \bibinfo {author} {\bibfnamefont {T.}~\bibnamefont {Tokushima}}, \bibinfo
  {author} {\bibfnamefont {D.}~\bibnamefont {Nordlund}}, \bibinfo {author}
  {\bibfnamefont {Y.}~\bibnamefont {Harada}}, \bibinfo {author} {\bibfnamefont
  {U.}~\bibnamefont {Bergmann}}, \bibinfo {author} {\bibfnamefont
  {M.}~\bibnamefont {Niebuhr}}, \bibinfo {author} {\bibfnamefont {T.~M.}\
  \bibnamefont {Weiss}}, \bibinfo {author} {\bibfnamefont {Y.}~\bibnamefont
  {Horikawa}}, \bibinfo {author} {\bibfnamefont {M.}~\bibnamefont {Leetmaa}},
  \bibinfo {author} {\bibfnamefont {M.~P.}\ \bibnamefont {Ljungberg}}, \bibinfo
  {author} {\bibfnamefont {O.}~\bibnamefont {Takahashi}}, \bibinfo {author}
  {\bibfnamefont {A.}~\bibnamefont {Lenz}}, \bibinfo {author} {\bibfnamefont
  {L.}~\bibnamefont {Ojam\"{a}e}}, \bibinfo {author} {\bibfnamefont {A.~P.}\
  \bibnamefont {Lyubartsev}}, \bibinfo {author} {\bibfnamefont
  {S.}~\bibnamefont {Shin}}, \bibinfo {author} {\bibfnamefont {L.~G.~M.}\
  \bibnamefont {Pettersson}}, \ and\ \bibinfo {author} {\bibfnamefont
  {A.}~\bibnamefont {Nilsson}},\ }\href@noop {} {\bibfield  {journal} {\bibinfo
   {journal} {Proc. Natl. Acad. Sci. USA}\ }\textbf {\bibinfo {volume} {106}},\
  \bibinfo {pages} {15214} (\bibinfo {year} {2009})}\BibitemShut {NoStop}%
\bibitem [{\citenamefont {Starr}, \citenamefont {Sciortino},\ and\
  \citenamefont {Stanley}(1999)}]{starr_1999}%
  \BibitemOpen
  \bibfield  {author} {\bibinfo {author} {\bibfnamefont {F.~W.}\ \bibnamefont
  {Starr}}, \bibinfo {author} {\bibfnamefont {F.}~\bibnamefont {Sciortino}}, \
  and\ \bibinfo {author} {\bibfnamefont {H.~E.}\ \bibnamefont {Stanley}},\
  }\href@noop {} {\bibfield  {journal} {\bibinfo  {journal} {Phys. Rev. E}\
  }\textbf {\bibinfo {volume} {60}},\ \bibinfo {pages} {6757} (\bibinfo {year}
  {1999})}\BibitemShut {NoStop}%
\bibitem [{\citenamefont {Faraone}\ \emph {et~al.}(2004)\citenamefont
  {Faraone}, \citenamefont {Liu}, \citenamefont {Mou}, \citenamefont {Yen},\
  and\ \citenamefont {Chen}}]{faraone_2004}%
  \BibitemOpen
  \bibfield  {author} {\bibinfo {author} {\bibfnamefont {A.}~\bibnamefont
  {Faraone}}, \bibinfo {author} {\bibfnamefont {L.}~\bibnamefont {Liu}},
  \bibinfo {author} {\bibfnamefont {C.~Y.}\ \bibnamefont {Mou}}, \bibinfo
  {author} {\bibfnamefont {C.~W.}\ \bibnamefont {Yen}}, \ and\ \bibinfo
  {author} {\bibfnamefont {S.-H.}\ \bibnamefont {Chen}},\ }\href@noop {}
  {\bibfield  {journal} {\bibinfo  {journal} {J. Chem. Phys.}\ }\textbf
  {\bibinfo {volume} {121}},\ \bibinfo {pages} {10843} (\bibinfo {year}
  {2004})}\BibitemShut {NoStop}%
\bibitem [{\citenamefont {Liu}\ \emph {et~al.}(2005)\citenamefont {Liu},
  \citenamefont {Chen}, \citenamefont {Faraone}, \citenamefont {Yen},\ and\
  \citenamefont {Mou}}]{liu_2005}%
  \BibitemOpen
  \bibfield  {author} {\bibinfo {author} {\bibfnamefont {L.}~\bibnamefont
  {Liu}}, \bibinfo {author} {\bibfnamefont {S.-H.}\ \bibnamefont {Chen}},
  \bibinfo {author} {\bibfnamefont {A.}~\bibnamefont {Faraone}}, \bibinfo
  {author} {\bibfnamefont {C.~W.}\ \bibnamefont {Yen}}, \ and\ \bibinfo
  {author} {\bibfnamefont {C.~Y.}\ \bibnamefont {Mou}},\ }\href@noop {}
  {\bibfield  {journal} {\bibinfo  {journal} {Phys. Rev. Lett.}\ }\textbf
  {\bibinfo {volume} {95}},\ \bibinfo {pages} {117802} (\bibinfo {year}
  {2005})}\BibitemShut {NoStop}%
\bibitem [{\citenamefont {Mallamace}\ \emph {et~al.}(2006)\citenamefont
  {Mallamace}, \citenamefont {Broccio}, \citenamefont {Corsaro}, \citenamefont
  {Faraone}, \citenamefont {Wanderlingh}, \citenamefont {Liu}, \citenamefont
  {Mou},\ and\ \citenamefont {Chen}}]{mallamace_2006}%
  \BibitemOpen
  \bibfield  {author} {\bibinfo {author} {\bibfnamefont {F.}~\bibnamefont
  {Mallamace}}, \bibinfo {author} {\bibfnamefont {M.}~\bibnamefont {Broccio}},
  \bibinfo {author} {\bibfnamefont {C.}~\bibnamefont {Corsaro}}, \bibinfo
  {author} {\bibfnamefont {A.}~\bibnamefont {Faraone}}, \bibinfo {author}
  {\bibfnamefont {U.}~\bibnamefont {Wanderlingh}}, \bibinfo {author}
  {\bibfnamefont {L.}~\bibnamefont {Liu}}, \bibinfo {author} {\bibfnamefont
  {C.~Y.}\ \bibnamefont {Mou}}, \ and\ \bibinfo {author} {\bibfnamefont
  {S.-H.}\ \bibnamefont {Chen}},\ }\href@noop {} {\bibfield  {journal}
  {\bibinfo  {journal} {J. Chem. Phys.}\ }\textbf {\bibinfo {volume} {124}},\
  \bibinfo {pages} {161102} (\bibinfo {year} {2006})}\BibitemShut {NoStop}%
\bibitem [{\citenamefont {Zheng}\ \emph {et~al.}(2009)\citenamefont {Zheng},
  \citenamefont {Lagi}, \citenamefont {Fratini}, \citenamefont {Baglioni},
  \citenamefont {Mamontov},\ and\ \citenamefont {Chen}}]{zhang_2009}%
  \BibitemOpen
  \bibfield  {author} {\bibinfo {author} {\bibfnamefont {Y.}~\bibnamefont
  {Zheng}}, \bibinfo {author} {\bibfnamefont {M.}~\bibnamefont {Lagi}},
  \bibinfo {author} {\bibfnamefont {E.}~\bibnamefont {Fratini}}, \bibinfo
  {author} {\bibfnamefont {P.}~\bibnamefont {Baglioni}}, \bibinfo {author}
  {\bibfnamefont {E.}~\bibnamefont {Mamontov}}, \ and\ \bibinfo {author}
  {\bibfnamefont {S.-H.}\ \bibnamefont {Chen}},\ }\href@noop {} {\bibfield
  {journal} {\bibinfo  {journal} {Phys. Rev. E}\ }\textbf {\bibinfo {volume}
  {79}},\ \bibinfo {pages} {040201} (\bibinfo {year} {2009})}\BibitemShut
  {NoStop}%
\bibitem [{\citenamefont {Gallo}, \citenamefont {Rovere},\ and\ \citenamefont
  {Chen}(2010)}]{gallo_2010}%
  \BibitemOpen
  \bibfield  {author} {\bibinfo {author} {\bibfnamefont {P.}~\bibnamefont
  {Gallo}}, \bibinfo {author} {\bibfnamefont {M.}~\bibnamefont {Rovere}}, \
  and\ \bibinfo {author} {\bibfnamefont {S.-H.}\ \bibnamefont {Chen}},\
  }\href@noop {} {\bibfield  {journal} {\bibinfo  {journal} {J. Phys. Chem.
  Lett.}\ }\textbf {\bibinfo {volume} {1}},\ \bibinfo {pages} {729} (\bibinfo
  {year} {2010})}\BibitemShut {NoStop}%
\bibitem [{\citenamefont {Gallo}\ and\ \citenamefont
  {Rovere}(2012)}]{gallo_2012}%
  \BibitemOpen
  \bibfield  {author} {\bibinfo {author} {\bibfnamefont {P.}~\bibnamefont
  {Gallo}}\ and\ \bibinfo {author} {\bibfnamefont {M.}~\bibnamefont {Rovere}},\
  }\href@noop {} {\bibfield  {journal} {\bibinfo  {journal} {J. Chem. Phys.}\
  }\textbf {\bibinfo {volume} {137}},\ \bibinfo {pages} {164503} (\bibinfo
  {year} {2012})}\BibitemShut {NoStop}%
\bibitem [{\citenamefont {Gallo}, \citenamefont {Rovere},\ and\ \citenamefont
  {Chen}(tter)}]{gallo_2012_2}%
  \BibitemOpen
  \bibfield  {author} {\bibinfo {author} {\bibfnamefont {P.}~\bibnamefont
  {Gallo}}, \bibinfo {author} {\bibfnamefont {M.}~\bibnamefont {Rovere}}, \
  and\ \bibinfo {author} {\bibfnamefont {S.~H.}\ \bibnamefont {Chen}},\
  }\href@noop {} {\bibfield  {journal} {\bibinfo  {journal} {2012}\ }\textbf
  {\bibinfo {volume} {24}},\ \bibinfo {pages} {064109} (\bibinfo {year} {J.
  Phys.: Condens. Matter})}\BibitemShut {NoStop}%
\bibitem [{\citenamefont {Wang}\ \emph {et~al.}(2015)\citenamefont {Wang},
  \citenamefont {Le}, \citenamefont {Ito}, \citenamefont {ao}, \citenamefont
  {Tyagi},\ and\ \citenamefont {Chen}}]{wang_2015}%
  \BibitemOpen
  \bibfield  {author} {\bibinfo {author} {\bibfnamefont {Z.}~\bibnamefont
  {Wang}}, \bibinfo {author} {\bibfnamefont {P.}~\bibnamefont {Le}}, \bibinfo
  {author} {\bibfnamefont {K.}~\bibnamefont {Ito}}, \bibinfo {author}
  {\bibfnamefont {J.~B.~L.}\ \bibnamefont {ao}}, \bibinfo {author}
  {\bibfnamefont {M.}~\bibnamefont {Tyagi}}, \ and\ \bibinfo {author}
  {\bibfnamefont {S.-H.}\ \bibnamefont {Chen}},\ }\href@noop {} {\bibfield
  {journal} {\bibinfo  {journal} {J. Chem. Phys.}\ }\textbf {\bibinfo {volume}
  {143}},\ \bibinfo {pages} {114508} (\bibinfo {year} {2015})}\BibitemShut
  {NoStop}%
\bibitem [{\citenamefont {Xu}\ \emph {et~al.}(2016)\citenamefont {Xu},
  \citenamefont {Petrik}, \citenamefont {Smith}, \citenamefont {Kay},\ and\
  \citenamefont {Kimmel}}]{YXu_2016}%
  \BibitemOpen
  \bibfield  {author} {\bibinfo {author} {\bibfnamefont {Y.}~\bibnamefont
  {Xu}}, \bibinfo {author} {\bibfnamefont {N.~G.}\ \bibnamefont {Petrik}},
  \bibinfo {author} {\bibfnamefont {R.~S.}\ \bibnamefont {Smith}}, \bibinfo
  {author} {\bibfnamefont {B.~D.}\ \bibnamefont {Kay}}, \ and\ \bibinfo
  {author} {\bibfnamefont {G.~A.}\ \bibnamefont {Kimmel}},\ }\href@noop {}
  {\bibfield  {journal} {\bibinfo  {journal} {Proc. Natl. Acad. Sci. USA}\
  }\textbf {\bibinfo {volume} {113}},\ \bibinfo {pages} {14921} (\bibinfo
  {year} {2016})}\BibitemShut {NoStop}%
\bibitem [{\citenamefont {{De Marzio}}\ \emph {et~al.}(2016)\citenamefont {{De
  Marzio}}, \citenamefont {Camicasca}, \citenamefont {Rovere},\ and\
  \citenamefont {Gallo}}]{demarzio_2016}%
  \BibitemOpen
  \bibfield  {author} {\bibinfo {author} {\bibfnamefont {M.}~\bibnamefont {{De
  Marzio}}}, \bibinfo {author} {\bibfnamefont {G.}~\bibnamefont {Camicasca}},
  \bibinfo {author} {\bibfnamefont {M.}~\bibnamefont {Rovere}}, \ and\ \bibinfo
  {author} {\bibfnamefont {P.}~\bibnamefont {Gallo}},\ }\href@noop {}
  {\bibfield  {journal} {\bibinfo  {journal} {J. Chem. Phys.}\ }\textbf
  {\bibinfo {volume} {114}},\ \bibinfo {pages} {074503} (\bibinfo {year}
  {2016})}\BibitemShut {NoStop}%
\bibitem [{\citenamefont {{De Marzio}}\ \emph {et~al.}(2017)\citenamefont {{De
  Marzio}}, \citenamefont {Camicasca}, \citenamefont {Rovere},\ and\
  \citenamefont {Gallo}}]{demarzio_2017}%
  \BibitemOpen
  \bibfield  {author} {\bibinfo {author} {\bibfnamefont {M.}~\bibnamefont {{De
  Marzio}}}, \bibinfo {author} {\bibfnamefont {G.}~\bibnamefont {Camicasca}},
  \bibinfo {author} {\bibfnamefont {M.}~\bibnamefont {Rovere}}, \ and\ \bibinfo
  {author} {\bibfnamefont {P.}~\bibnamefont {Gallo}},\ }\href@noop {}
  {\bibfield  {journal} {\bibinfo  {journal} {J. Chem. Phys.}\ }\textbf
  {\bibinfo {volume} {146}},\ \bibinfo {pages} {084502} (\bibinfo {year}
  {2017})}\BibitemShut {NoStop}%
\bibitem [{\citenamefont {Shi}, \citenamefont {Russo},\ and\ \citenamefont
  {Tanaka}(2018{\natexlab{a}})}]{shi_2018_1}%
  \BibitemOpen
  \bibfield  {author} {\bibinfo {author} {\bibfnamefont {R.}~\bibnamefont
  {Shi}}, \bibinfo {author} {\bibfnamefont {J.}~\bibnamefont {Russo}}, \ and\
  \bibinfo {author} {\bibfnamefont {H.}~\bibnamefont {Tanaka}},\ }\href@noop {}
  {\bibfield  {journal} {\bibinfo  {journal} {Proc. Natl. Acad. Sci. USA}\
  }\textbf {\bibinfo {volume} {115}},\ \bibinfo {pages} {9444} (\bibinfo {year}
  {2018}{\natexlab{a}})}\BibitemShut {NoStop}%
\bibitem [{\citenamefont {Shi}, \citenamefont {Russo},\ and\ \citenamefont
  {Tanaka}(2018{\natexlab{b}})}]{shi_2018_2}%
  \BibitemOpen
  \bibfield  {author} {\bibinfo {author} {\bibfnamefont {R.}~\bibnamefont
  {Shi}}, \bibinfo {author} {\bibfnamefont {J.}~\bibnamefont {Russo}}, \ and\
  \bibinfo {author} {\bibfnamefont {H.}~\bibnamefont {Tanaka}},\ }\href@noop {}
  {\bibfield  {journal} {\bibinfo  {journal} {J. Chem. Phys.}\ }\textbf
  {\bibinfo {volume} {149}},\ \bibinfo {pages} {224502} (\bibinfo {year}
  {2018}{\natexlab{b}})}\BibitemShut {NoStop}%
\bibitem [{\citenamefont {Eisenberg}\ and\ \citenamefont
  {Kauzmann}(1969)}]{kauzmann}%
  \BibitemOpen
  \bibfield  {author} {\bibinfo {author} {\bibfnamefont {D.}~\bibnamefont
  {Eisenberg}}\ and\ \bibinfo {author} {\bibfnamefont {W.}~\bibnamefont
  {Kauzmann}},\ }\href@noop {} {\emph {\bibinfo {title} {The Structure and
  Properties of Water}}}\ (\bibinfo  {publisher} {Oxford Univ. Press},\
  \bibinfo {year} {1969})\BibitemShut {NoStop}%
\bibitem [{\citenamefont {Angell}(1995)}]{angell_1995}%
  \BibitemOpen
  \bibfield  {author} {\bibinfo {author} {\bibfnamefont {C.~A.}\ \bibnamefont
  {Angell}},\ }\href@noop {} {\bibfield  {journal} {\bibinfo  {journal}
  {Science}\ }\textbf {\bibinfo {volume} {267}},\ \bibinfo {pages} {1924}
  (\bibinfo {year} {1995})}\BibitemShut {NoStop}%
\bibitem [{\citenamefont {Debenedetti}(2003)}]{debenedetti_2003}%
  \BibitemOpen
  \bibfield  {author} {\bibinfo {author} {\bibfnamefont {P.~G.}\ \bibnamefont
  {Debenedetti}},\ }\href@noop {} {\bibfield  {journal} {\bibinfo  {journal}
  {J. Phys.: Condens. Matter}\ }\textbf {\bibinfo {volume} {15}},\ \bibinfo
  {pages} {R1669} (\bibinfo {year} {2003})}\BibitemShut {NoStop}%
\bibitem [{Note1()}]{Note1}%
  \BibitemOpen
  \bibinfo {note} {The Widom line can also be defined, e.g., on the isothermal
  pathway~\cite {fomin_2015,schienbein_2018}}\BibitemShut {NoStop}%
\bibitem [{\citenamefont {Abascal}\ and\ \citenamefont
  {Vega}(2005)}]{tip4p2005}%
  \BibitemOpen
  \bibfield  {author} {\bibinfo {author} {\bibfnamefont {J.~L.~F.}\
  \bibnamefont {Abascal}}\ and\ \bibinfo {author} {\bibfnamefont
  {C.}~\bibnamefont {Vega}},\ }\href@noop {} {\bibfield  {journal} {\bibinfo
  {journal} {J. Chem. Phys.}\ }\textbf {\bibinfo {volume} {123}},\ \bibinfo
  {pages} {234505} (\bibinfo {year} {2005})}\BibitemShut {NoStop}%
\bibitem [{\citenamefont {Nos/'e}(1984)}]{nose}%
  \BibitemOpen
  \bibfield  {author} {\bibinfo {author} {\bibfnamefont {S.}~\bibnamefont
  {Nos/'e}},\ }\href@noop {} {\bibfield  {journal} {\bibinfo  {journal} {Mol.
  Phys.}\ }\textbf {\bibinfo {volume} {52}},\ \bibinfo {pages} {255} (\bibinfo
  {year} {1984})}\BibitemShut {NoStop}%
\bibitem [{\citenamefont {Hoover}(1985)}]{hoover}%
  \BibitemOpen
  \bibfield  {author} {\bibinfo {author} {\bibfnamefont {W.~G.}\ \bibnamefont
  {Hoover}},\ }\href@noop {} {\bibfield  {journal} {\bibinfo  {journal} {Phys.
  Rev. A}\ }\textbf {\bibinfo {volume} {31}},\ \bibinfo {pages} {1695}
  (\bibinfo {year} {1985})}\BibitemShut {NoStop}%
\bibitem [{\citenamefont {Parrinello}\ and\ \citenamefont
  {Rahman}(1981)}]{parrinello_rahman}%
  \BibitemOpen
  \bibfield  {author} {\bibinfo {author} {\bibfnamefont {M.}~\bibnamefont
  {Parrinello}}\ and\ \bibinfo {author} {\bibfnamefont {A.}~\bibnamefont
  {Rahman}},\ }\href@noop {} {\bibfield  {journal} {\bibinfo  {journal} {J.
  Appl. Phys.}\ }\textbf {\bibinfo {volume} {52}},\ \bibinfo {pages} {7182}
  (\bibinfo {year} {1981})}\BibitemShut {NoStop}%
\bibitem [{\citenamefont {Pronk}\ \emph {et~al.}(2013)\citenamefont {Pronk},
  \citenamefont {P\`all}, \citenamefont {Schulz}, \citenamefont {Larsson},
  \citenamefont {Bjelkmar}, \citenamefont {Apostolov}, \citenamefont {Shirts},
  \citenamefont {Smith}, \citenamefont {Kasson}, \citenamefont {van~der Spoel},
  \citenamefont {Hess},\ and\ \citenamefont {Lindahl}}]{gromacs}%
  \BibitemOpen
  \bibfield  {author} {\bibinfo {author} {\bibfnamefont {S.}~\bibnamefont
  {Pronk}}, \bibinfo {author} {\bibfnamefont {S.}~\bibnamefont {P\`all}},
  \bibinfo {author} {\bibfnamefont {R.}~\bibnamefont {Schulz}}, \bibinfo
  {author} {\bibfnamefont {P.}~\bibnamefont {Larsson}}, \bibinfo {author}
  {\bibfnamefont {P.}~\bibnamefont {Bjelkmar}}, \bibinfo {author}
  {\bibfnamefont {R.}~\bibnamefont {Apostolov}}, \bibinfo {author}
  {\bibfnamefont {M.~R.}\ \bibnamefont {Shirts}}, \bibinfo {author}
  {\bibfnamefont {J.~C.}\ \bibnamefont {Smith}}, \bibinfo {author}
  {\bibfnamefont {P.~M.}\ \bibnamefont {Kasson}}, \bibinfo {author}
  {\bibfnamefont {D.}~\bibnamefont {van~der Spoel}}, \bibinfo {author}
  {\bibfnamefont {B.}~\bibnamefont {Hess}}, \ and\ \bibinfo {author}
  {\bibfnamefont {E.}~\bibnamefont {Lindahl}},\ }\href@noop {} {\bibfield
  {journal} {\bibinfo  {journal} {Bioinformatics}\ }\textbf {\bibinfo {volume}
  {29}},\ \bibinfo {pages} {845} (\bibinfo {year} {2013})}\BibitemShut
  {NoStop}%
\bibitem [{\citenamefont {Hansen}\ and\ \citenamefont
  {McDonald}(2006)}]{hansen}%
  \BibitemOpen
  \bibfield  {author} {\bibinfo {author} {\bibfnamefont {J.-P.}\ \bibnamefont
  {Hansen}}\ and\ \bibinfo {author} {\bibfnamefont {I.~R.}\ \bibnamefont
  {McDonald}},\ }\href@noop {} {\emph {\bibinfo {title} {Theory of Simple
  Liquids}}},\ \bibinfo {edition} {3rd}\ ed.\ (\bibinfo  {publisher}
  {Elsevier},\ \bibinfo {year} {2006})\BibitemShut {NoStop}%
\bibitem [{\citenamefont {Martelli}\ \emph {et~al.}(2016)\citenamefont
  {Martelli}, \citenamefont {Ko}, \citenamefont {O\u{g}uz},\ and\ \citenamefont
  {Car}}]{martelli_LOM}%
  \BibitemOpen
  \bibfield  {author} {\bibinfo {author} {\bibfnamefont {F.}~\bibnamefont
  {Martelli}}, \bibinfo {author} {\bibfnamefont {H.-Y.}\ \bibnamefont {Ko}},
  \bibinfo {author} {\bibfnamefont {E.~C.}\ \bibnamefont {O\u{g}uz}}, \ and\
  \bibinfo {author} {\bibfnamefont {R.}~\bibnamefont {Car}},\ }\href@noop {}
  {\bibfield  {journal} {\bibinfo  {journal} {Phys. Rev. B}\ }\textbf {\bibinfo
  {volume} {97}},\ \bibinfo {pages} {064105} (\bibinfo {year}
  {2016})}\BibitemShut {NoStop}%
\bibitem [{\citenamefont {Martelli}\ \emph
  {et~al.}(2018{\natexlab{b}})\citenamefont {Martelli}, \citenamefont {Ko},
  \citenamefont {Borallo},\ and\ \citenamefont {Franzese}}]{martelli_fop}%
  \BibitemOpen
  \bibfield  {author} {\bibinfo {author} {\bibfnamefont {F.}~\bibnamefont
  {Martelli}}, \bibinfo {author} {\bibfnamefont {H.-Y.}\ \bibnamefont {Ko}},
  \bibinfo {author} {\bibfnamefont {C.~C.}\ \bibnamefont {Borallo}}, \ and\
  \bibinfo {author} {\bibfnamefont {G.}~\bibnamefont {Franzese}},\ }\href@noop
  {} {\bibfield  {journal} {\bibinfo  {journal} {Front. Phys.}\ }\textbf
  {\bibinfo {volume} {13}},\ \bibinfo {pages} {136801} (\bibinfo {year}
  {2018}{\natexlab{b}})}\BibitemShut {NoStop}%
\bibitem [{\citenamefont {Samatas}\ \emph {et~al.}(2018)\citenamefont
  {Samatas}, \citenamefont {Calero}, \citenamefont {Martelli},\ and\
  \citenamefont {Franzese}}]{samatas_2018}%
  \BibitemOpen
  \bibfield  {author} {\bibinfo {author} {\bibfnamefont {S.}~\bibnamefont
  {Samatas}}, \bibinfo {author} {\bibfnamefont {C.}~\bibnamefont {Calero}},
  \bibinfo {author} {\bibfnamefont {F.}~\bibnamefont {Martelli}}, \ and\
  \bibinfo {author} {\bibfnamefont {G.}~\bibnamefont {Franzese}},\ }\href@noop
  {} {\bibfield  {journal} {\bibinfo  {journal} {arXiv:1811.01911
  [cond-mat.soft]}\ } (\bibinfo {year} {2018})}\BibitemShut {NoStop}%
\bibitem [{\citenamefont {Santra}\ \emph {et~al.}(2018)\citenamefont {Santra},
  \citenamefont {Ko}, \citenamefont {Yeh}, \citenamefont {Martelli},
  \citenamefont {Kaganovich}, \citenamefont {Raitses},\ and\ \citenamefont
  {Car}}]{santra_bnnt}%
  \BibitemOpen
  \bibfield  {author} {\bibinfo {author} {\bibfnamefont {B.}~\bibnamefont
  {Santra}}, \bibinfo {author} {\bibfnamefont {H.-Y.}\ \bibnamefont {Ko}},
  \bibinfo {author} {\bibfnamefont {Y.-W.}\ \bibnamefont {Yeh}}, \bibinfo
  {author} {\bibfnamefont {F.}~\bibnamefont {Martelli}}, \bibinfo {author}
  {\bibfnamefont {I.}~\bibnamefont {Kaganovich}}, \bibinfo {author}
  {\bibfnamefont {Y.}~\bibnamefont {Raitses}}, \ and\ \bibinfo {author}
  {\bibfnamefont {R.}~\bibnamefont {Car}},\ }\href@noop {} {\bibfield
  {journal} {\bibinfo  {journal} {Nanoscale}\ }\textbf {\bibinfo {volume}
  {10}},\ \bibinfo {pages} {22223} (\bibinfo {year} {2018})}\BibitemShut
  {NoStop}%
\bibitem [{\citenamefont {Shiratani}\ and\ \citenamefont
  {Sasai}(1996)}]{LSI_1}%
  \BibitemOpen
  \bibfield  {author} {\bibinfo {author} {\bibfnamefont {E.}~\bibnamefont
  {Shiratani}}\ and\ \bibinfo {author} {\bibfnamefont {M.}~\bibnamefont
  {Sasai}},\ }\href@noop {} {\bibfield  {journal} {\bibinfo  {journal} {J.
  Chem. Phys.}\ }\textbf {\bibinfo {volume} {104}},\ \bibinfo {pages} {7671}
  (\bibinfo {year} {1996})}\BibitemShut {NoStop}%
\bibitem [{\citenamefont {Shiratani}\ and\ \citenamefont
  {Sasai}(1998)}]{LSI_2}%
  \BibitemOpen
  \bibfield  {author} {\bibinfo {author} {\bibfnamefont {E.}~\bibnamefont
  {Shiratani}}\ and\ \bibinfo {author} {\bibfnamefont {M.}~\bibnamefont
  {Sasai}},\ }\href@noop {} {\bibfield  {journal} {\bibinfo  {journal} {J.
  Chem. Phys.}\ }\textbf {\bibinfo {volume} {108}},\ \bibinfo {pages} {3264}
  (\bibinfo {year} {1998})}\BibitemShut {NoStop}%
\bibitem [{\citenamefont {Soper}\ and\ \citenamefont
  {Ricci}(2000)}]{soper_2000}%
  \BibitemOpen
  \bibfield  {author} {\bibinfo {author} {\bibfnamefont {A.~K.}\ \bibnamefont
  {Soper}}\ and\ \bibinfo {author} {\bibfnamefont {M.~A.}\ \bibnamefont
  {Ricci}},\ }\href@noop {} {\bibfield  {journal} {\bibinfo  {journal} {Phys.
  Rev. Lett.}\ }\textbf {\bibinfo {volume} {84}},\ \bibinfo {pages} {2881}
  (\bibinfo {year} {2000})}\BibitemShut {NoStop}%
\bibitem [{\citenamefont {Svishchev}\ and\ \citenamefont
  {Kusalik}(1993)}]{svishchev_1993}%
  \BibitemOpen
  \bibfield  {author} {\bibinfo {author} {\bibfnamefont {I.~M.}\ \bibnamefont
  {Svishchev}}\ and\ \bibinfo {author} {\bibfnamefont {P.~G.}\ \bibnamefont
  {Kusalik}},\ }\href@noop {} {\bibfield  {journal} {\bibinfo  {journal} {J.
  Chem. Phys.}\ }\textbf {\bibinfo {volume} {99}},\ \bibinfo {pages} {3049}
  (\bibinfo {year} {1993})}\BibitemShut {NoStop}%
\bibitem [{\citenamefont {Schwegler}, \citenamefont {Galli},\ and\
  \citenamefont {Gygi}(2000)}]{schwegler_2000}%
  \BibitemOpen
  \bibfield  {author} {\bibinfo {author} {\bibfnamefont {E.}~\bibnamefont
  {Schwegler}}, \bibinfo {author} {\bibfnamefont {G.}~\bibnamefont {Galli}}, \
  and\ \bibinfo {author} {\bibfnamefont {F.}~\bibnamefont {Gygi}},\ }\href@noop
  {} {\bibfield  {journal} {\bibinfo  {journal} {Phys. Rev. Lett.}\ }\textbf
  {\bibinfo {volume} {84}},\ \bibinfo {pages} {2429} (\bibinfo {year}
  {2000})}\BibitemShut {NoStop}%
\bibitem [{\citenamefont {Saitta}\ and\ \citenamefont
  {Datchi}(2003)}]{saitta_2003}%
  \BibitemOpen
  \bibfield  {author} {\bibinfo {author} {\bibfnamefont {A.~M.}\ \bibnamefont
  {Saitta}}\ and\ \bibinfo {author} {\bibfnamefont {F.}~\bibnamefont
  {Datchi}},\ }\href@noop {} {\bibfield  {journal} {\bibinfo  {journal} {Phys.
  Rev. E}\ }\textbf {\bibinfo {volume} {67}},\ \bibinfo {pages} {020201(R)}
  (\bibinfo {year} {2003})}\BibitemShut {NoStop}%
\bibitem [{\citenamefont {Sciortino}, \citenamefont {Geiger},\ and\
  \citenamefont {Stanley}(1990)}]{sciortino_1990}%
  \BibitemOpen
  \bibfield  {author} {\bibinfo {author} {\bibfnamefont {F.}~\bibnamefont
  {Sciortino}}, \bibinfo {author} {\bibfnamefont {A.}~\bibnamefont {Geiger}}, \
  and\ \bibinfo {author} {\bibfnamefont {H.~E.}\ \bibnamefont {Stanley}},\
  }\href@noop {} {\bibfield  {journal} {\bibinfo  {journal} {Phys. Rev. Lett.}\
  }\textbf {\bibinfo {volume} {65}},\ \bibinfo {pages} {3452} (\bibinfo {year}
  {1990})}\BibitemShut {NoStop}%
\bibitem [{\citenamefont {Sciortino}, \citenamefont {Geiger},\ and\
  \citenamefont {Stanley}(1991)}]{sciortino_1991}%
  \BibitemOpen
  \bibfield  {author} {\bibinfo {author} {\bibfnamefont {F.}~\bibnamefont
  {Sciortino}}, \bibinfo {author} {\bibfnamefont {A.}~\bibnamefont {Geiger}}, \
  and\ \bibinfo {author} {\bibfnamefont {H.~E.}\ \bibnamefont {Stanley}},\
  }\href@noop {} {\bibfield  {journal} {\bibinfo  {journal} {Nature}\ }\textbf
  {\bibinfo {volume} {354}},\ \bibinfo {pages} {218} (\bibinfo {year}
  {1991})}\BibitemShut {NoStop}%
\bibitem [{\citenamefont {Kumar}\ \emph {et~al.}(2006)\citenamefont {Kumar},
  \citenamefont {Franzese}, \citenamefont {Buldyrev},\ and\ \citenamefont
  {Stanley}}]{kumar_2006}%
  \BibitemOpen
  \bibfield  {author} {\bibinfo {author} {\bibfnamefont {P.}~\bibnamefont
  {Kumar}}, \bibinfo {author} {\bibfnamefont {G.}~\bibnamefont {Franzese}},
  \bibinfo {author} {\bibfnamefont {S.~V.}\ \bibnamefont {Buldyrev}}, \ and\
  \bibinfo {author} {\bibfnamefont {H.~E.}\ \bibnamefont {Stanley}},\
  }\href@noop {} {\bibfield  {journal} {\bibinfo  {journal} {Phys. Rev. E}\
  }\textbf {\bibinfo {volume} {73}},\ \bibinfo {pages} {041505} (\bibinfo
  {year} {2006})}\BibitemShut {NoStop}%
\bibitem [{\citenamefont {Franzese}\ and\ \citenamefont
  {Stanley}(2007)}]{franzese_2007}%
  \BibitemOpen
  \bibfield  {author} {\bibinfo {author} {\bibfnamefont {G.}~\bibnamefont
  {Franzese}}\ and\ \bibinfo {author} {\bibfnamefont {H.~E.}\ \bibnamefont
  {Stanley}},\ }\href@noop {} {\bibfield  {journal} {\bibinfo  {journal} {J.
  Phys.: Condens. Matter}\ }\textbf {\bibinfo {volume} {19}},\ \bibinfo {pages}
  {205126} (\bibinfo {year} {2007})}\BibitemShut {NoStop}%
\bibitem [{\citenamefont {Lapini}\ \emph {et~al.}(2016)\citenamefont {Lapini},
  \citenamefont {Pagliai}, \citenamefont {Fanetti}, \citenamefont {Citroni},
  \citenamefont {Scandolo}, \citenamefont {Bini},\ and\ \citenamefont
  {Righini}}]{lapini_2016}%
  \BibitemOpen
  \bibfield  {author} {\bibinfo {author} {\bibfnamefont {A.}~\bibnamefont
  {Lapini}}, \bibinfo {author} {\bibfnamefont {M.}~\bibnamefont {Pagliai}},
  \bibinfo {author} {\bibfnamefont {S.}~\bibnamefont {Fanetti}}, \bibinfo
  {author} {\bibfnamefont {M.}~\bibnamefont {Citroni}}, \bibinfo {author}
  {\bibfnamefont {S.}~\bibnamefont {Scandolo}}, \bibinfo {author}
  {\bibfnamefont {R.}~\bibnamefont {Bini}}, \ and\ \bibinfo {author}
  {\bibfnamefont {R.}~\bibnamefont {Righini}},\ }\href@noop {} {\bibfield
  {journal} {\bibinfo  {journal} {J. Phys. Chem. Lett.}\ }\textbf {\bibinfo
  {volume} {7}},\ \bibinfo {pages} {3579} (\bibinfo {year} {2016})}\BibitemShut
  {NoStop}%
\bibitem [{\citenamefont {Cuthbertson}\ and\ \citenamefont
  {Poole}(2011)}]{cuthbertson_2011}%
  \BibitemOpen
  \bibfield  {author} {\bibinfo {author} {\bibfnamefont {M.~J.}\ \bibnamefont
  {Cuthbertson}}\ and\ \bibinfo {author} {\bibfnamefont {P.~H.}\ \bibnamefont
  {Poole}},\ }\href@noop {} {\bibfield  {journal} {\bibinfo  {journal} {Phys.
  Rev. Lett.}\ }\textbf {\bibinfo {volume} {106}},\ \bibinfo {pages} {115706}
  (\bibinfo {year} {2011})}\BibitemShut {NoStop}%
\bibitem [{\citenamefont {Wikfeldt}, \citenamefont {Nilsson},\ and\
  \citenamefont {Pettersson}(2011)}]{wikfeldt_2011}%
  \BibitemOpen
  \bibfield  {author} {\bibinfo {author} {\bibfnamefont {K.~T.}\ \bibnamefont
  {Wikfeldt}}, \bibinfo {author} {\bibfnamefont {A.}~\bibnamefont {Nilsson}}, \
  and\ \bibinfo {author} {\bibfnamefont {L.~G.~M.}\ \bibnamefont
  {Pettersson}},\ }\href@noop {} {\bibfield  {journal} {\bibinfo  {journal}
  {Phys. Chem. Chem. Phys.}\ }\textbf {\bibinfo {volume} {13}},\ \bibinfo
  {pages} {19918} (\bibinfo {year} {2011})}\BibitemShut {NoStop}%
\bibitem [{\citenamefont {Appignanesi}, \citenamefont {Rodriguez},\ and\
  \citenamefont {Sciortino}(2009)}]{appignanesi_2009}%
  \BibitemOpen
  \bibfield  {author} {\bibinfo {author} {\bibfnamefont {G.~A.}\ \bibnamefont
  {Appignanesi}}, \bibinfo {author} {\bibfnamefont {J.~A.}\ \bibnamefont
  {Rodriguez}}, \ and\ \bibinfo {author} {\bibfnamefont {F.}~\bibnamefont
  {Sciortino}},\ }\href@noop {} {\bibfield  {journal} {\bibinfo  {journal}
  {Euro. Phys. J. E}\ }\textbf {\bibinfo {volume} {29}},\ \bibinfo {pages}
  {305} (\bibinfo {year} {2009})}\BibitemShut {NoStop}%
\bibitem [{\citenamefont {{Montes de Oca}}\ \emph {et~al.}(2016)\citenamefont
  {{Montes de Oca}}, \citenamefont {Rodriguez}, \citenamefont {Accordino},
  \citenamefont {Malaspina},\ and\ \citenamefont
  {Appignanesi}}]{montes_de_oca}%
  \BibitemOpen
  \bibfield  {author} {\bibinfo {author} {\bibfnamefont {J.~M.}\ \bibnamefont
  {{Montes de Oca}}}, \bibinfo {author} {\bibfnamefont {A.}~\bibnamefont
  {Rodriguez}}, \bibinfo {author} {\bibfnamefont {S.~R.}\ \bibnamefont
  {Accordino}}, \bibinfo {author} {\bibfnamefont {D.~C.}\ \bibnamefont
  {Malaspina}}, \ and\ \bibinfo {author} {\bibfnamefont {G.~A.}\ \bibnamefont
  {Appignanesi}},\ }\href@noop {} {\bibfield  {journal} {\bibinfo  {journal}
  {Eur. Phys. J. E}\ }\textbf {\bibinfo {volume} {39}},\ \bibinfo {pages} {124}
  (\bibinfo {year} {2016})}\BibitemShut {NoStop}%
\bibitem [{\citenamefont {Shi}\ and\ \citenamefont
  {Tanaka}(2018)}]{shi_2018_3}%
  \BibitemOpen
  \bibfield  {author} {\bibinfo {author} {\bibfnamefont {R.}~\bibnamefont
  {Shi}}\ and\ \bibinfo {author} {\bibfnamefont {H.}~\bibnamefont {Tanaka}},\
  }\href@noop {} {\bibfield  {journal} {\bibinfo  {journal} {J. Chem. Phys.}\
  }\textbf {\bibinfo {volume} {148}},\ \bibinfo {pages} {124503} (\bibinfo
  {year} {2018})}\BibitemShut {NoStop}%
\bibitem [{Note2()}]{Note2}%
  \BibitemOpen
  \bibinfo {note} {The position of the isosbestic point in the presence of
  thermal energy fluctuates within the range $0.12-0.14$ \r A$^2$. Any change
  within this range would produce very minor quantitative changes in our
  results. The position of the isosbestic point also depends on the cutoff
  employed in the evaluation of eq.~\ref {eq:Eq3}~\cite
  {accordino_2011}}\BibitemShut {NoStop}%
\bibitem [{\citenamefont {Saito}, \citenamefont {Bagchi},\ and\ \citenamefont
  {Ohmine}(2018)}]{saito_2018}%
  \BibitemOpen
  \bibfield  {author} {\bibinfo {author} {\bibfnamefont {S.}~\bibnamefont
  {Saito}}, \bibinfo {author} {\bibfnamefont {B.}~\bibnamefont {Bagchi}}, \
  and\ \bibinfo {author} {\bibfnamefont {I.}~\bibnamefont {Ohmine}},\
  }\href@noop {} {\bibfield  {journal} {\bibinfo  {journal} {J. Chem. Phys.}\
  }\textbf {\bibinfo {volume} {149}},\ \bibinfo {pages} {124504} (\bibinfo
  {year} {2018})}\BibitemShut {NoStop}%
\bibitem [{\citenamefont {Xu}\ \emph {et~al.}(2009)\citenamefont {Xu},
  \citenamefont {Mallamace}, \citenamefont {Yan}, \citenamefont {Starr},
  \citenamefont {Buldyrev},\ and\ \citenamefont {Stanley}}]{xu_2009}%
  \BibitemOpen
  \bibfield  {author} {\bibinfo {author} {\bibfnamefont {L.}~\bibnamefont
  {Xu}}, \bibinfo {author} {\bibfnamefont {F.}~\bibnamefont {Mallamace}},
  \bibinfo {author} {\bibfnamefont {Z.}~\bibnamefont {Yan}}, \bibinfo {author}
  {\bibfnamefont {F.~W.}\ \bibnamefont {Starr}}, \bibinfo {author}
  {\bibfnamefont {S.~V.}\ \bibnamefont {Buldyrev}}, \ and\ \bibinfo {author}
  {\bibfnamefont {H.~E.}\ \bibnamefont {Stanley}},\ }\href@noop {} {\bibfield
  {journal} {\bibinfo  {journal} {Nat. Phys.}\ }\textbf {\bibinfo {volume}
  {5}},\ \bibinfo {pages} {565} (\bibinfo {year} {2009})}\BibitemShut {NoStop}%
\bibitem [{\citenamefont {Mallamace}, \citenamefont {Corsaro},\ and\
  \citenamefont {Stanley}(2013)}]{mallamace_2013}%
  \BibitemOpen
  \bibfield  {author} {\bibinfo {author} {\bibfnamefont {F.}~\bibnamefont
  {Mallamace}}, \bibinfo {author} {\bibfnamefont {C.}~\bibnamefont {Corsaro}},
  \ and\ \bibinfo {author} {\bibfnamefont {H.~E.}\ \bibnamefont {Stanley}},\
  }\href@noop {} {\bibfield  {journal} {\bibinfo  {journal} {Proc. Natl. Acad.
  Sci. USA}\ }\textbf {\bibinfo {volume} {110}},\ \bibinfo {pages} {4899}
  (\bibinfo {year} {2013})}\BibitemShut {NoStop}%
\bibitem [{\citenamefont {Luzar}\ and\ \citenamefont
  {Chandler}(1996)}]{chandler_HB}%
  \BibitemOpen
  \bibfield  {author} {\bibinfo {author} {\bibfnamefont {A.}~\bibnamefont
  {Luzar}}\ and\ \bibinfo {author} {\bibfnamefont {D.}~\bibnamefont
  {Chandler}},\ }\href@noop {} {\bibfield  {journal} {\bibinfo  {journal}
  {Nature}\ }\textbf {\bibinfo {volume} {379}},\ \bibinfo {pages} {55}
  (\bibinfo {year} {1996})}\BibitemShut {NoStop}%
\bibitem [{\citenamefont {Prada-Gracia}, \citenamefont {Shevchuk},\ and\
  \citenamefont {Rao}(2013)}]{prada_2013}%
  \BibitemOpen
  \bibfield  {author} {\bibinfo {author} {\bibfnamefont {D.}~\bibnamefont
  {Prada-Gracia}}, \bibinfo {author} {\bibfnamefont {R.}~\bibnamefont
  {Shevchuk}}, \ and\ \bibinfo {author} {\bibfnamefont {F.}~\bibnamefont
  {Rao}},\ }\href@noop {} {\bibfield  {journal} {\bibinfo  {journal} {J. Chem.
  Phys.}\ }\textbf {\bibinfo {volume} {139}},\ \bibinfo {pages} {084501}
  (\bibinfo {year} {2013})}\BibitemShut {NoStop}%
\bibitem [{\citenamefont {Marto\u{n}\'{a}k}, \citenamefont {Donadio},\ and\
  \citenamefont {Parrinello}(2004)}]{martonak_2004}%
  \BibitemOpen
  \bibfield  {author} {\bibinfo {author} {\bibfnamefont {R.}~\bibnamefont
  {Marto\u{n}\'{a}k}}, \bibinfo {author} {\bibfnamefont {D.}~\bibnamefont
  {Donadio}}, \ and\ \bibinfo {author} {\bibfnamefont {M.}~\bibnamefont
  {Parrinello}},\ }\href@noop {} {\bibfield  {journal} {\bibinfo  {journal}
  {Phys. Rev. Lett.}\ }\textbf {\bibinfo {volume} {92}},\ \bibinfo {pages}
  {225702} (\bibinfo {year} {2004})}\BibitemShut {NoStop}%
\bibitem [{\citenamefont {Marto\u{n}\'{a}k}, \citenamefont {Donadio},\ and\
  \citenamefont {Parrinello}(2005)}]{martonak_2005}%
  \BibitemOpen
  \bibfield  {author} {\bibinfo {author} {\bibfnamefont {R.}~\bibnamefont
  {Marto\u{n}\'{a}k}}, \bibinfo {author} {\bibfnamefont {D.}~\bibnamefont
  {Donadio}}, \ and\ \bibinfo {author} {\bibfnamefont {M.}~\bibnamefont
  {Parrinello}},\ }\href@noop {} {\bibfield  {journal} {\bibinfo  {journal} {J.
  Chem. Phys.}\ }\textbf {\bibinfo {volume} {122}},\ \bibinfo {pages} {134501}
  (\bibinfo {year} {2005})}\BibitemShut {NoStop}%
\bibitem [{\citenamefont {Finney}\ \emph {et~al.}(2002)\citenamefont {Finney},
  \citenamefont {Hallbrucker}, \citenamefont {Kohl}, \citenamefont {Soper},\
  and\ \citenamefont {Bowron}}]{finney_2002}%
  \BibitemOpen
  \bibfield  {author} {\bibinfo {author} {\bibfnamefont {J.~L.}\ \bibnamefont
  {Finney}}, \bibinfo {author} {\bibfnamefont {A.}~\bibnamefont {Hallbrucker}},
  \bibinfo {author} {\bibfnamefont {I.}~\bibnamefont {Kohl}}, \bibinfo {author}
  {\bibfnamefont {A.~K.}\ \bibnamefont {Soper}}, \ and\ \bibinfo {author}
  {\bibfnamefont {D.~T.}\ \bibnamefont {Bowron}},\ }\href@noop {} {\bibfield
  {journal} {\bibinfo  {journal} {Phys. Rev. Lett.}\ }\textbf {\bibinfo
  {volume} {88}},\ \bibinfo {pages} {225503} (\bibinfo {year}
  {2002})}\BibitemShut {NoStop}%
\bibitem [{\citenamefont {Giovambattista}\ \emph {et~al.}(2016)\citenamefont
  {Giovambattista}, \citenamefont {Sciortino}, \citenamefont {Starr},\ and\
  \citenamefont {Poole}}]{giovambattista_2016}%
  \BibitemOpen
  \bibfield  {author} {\bibinfo {author} {\bibfnamefont {N.}~\bibnamefont
  {Giovambattista}}, \bibinfo {author} {\bibfnamefont {F.}~\bibnamefont
  {Sciortino}}, \bibinfo {author} {\bibfnamefont {F.~W.}\ \bibnamefont
  {Starr}}, \ and\ \bibinfo {author} {\bibfnamefont {P.~H.}\ \bibnamefont
  {Poole}},\ }\href@noop {} {\bibfield  {journal} {\bibinfo  {journal} {J.
  Chem. Phys.}\ }\textbf {\bibinfo {volume} {145}},\ \bibinfo {pages} {224501}
  (\bibinfo {year} {2016})}\BibitemShut {NoStop}%
\bibitem [{\citenamefont {Handle}\ and\ \citenamefont
  {Sciortino}(2018)}]{handle_2018}%
  \BibitemOpen
  \bibfield  {author} {\bibinfo {author} {\bibfnamefont {P.~H.}\ \bibnamefont
  {Handle}}\ and\ \bibinfo {author} {\bibfnamefont {F.}~\bibnamefont
  {Sciortino}},\ }\href@noop {} {\bibfield  {journal} {\bibinfo  {journal} {J.
  Chem. Phys.}\ }\textbf {\bibinfo {volume} {148}},\ \bibinfo {pages} {134505}
  (\bibinfo {year} {2018})}\BibitemShut {NoStop}%
\bibitem [{Note3()}]{Note3}%
  \BibitemOpen
  \bibinfo {note} {It is worthy to mention that in LDA, the interstitial
  molecules are not bonded to the central one~\cite
  {finney_2002,martelli_hyperuniformity}, suggesting that the HBN of LDA and
  LDL are substantially different, as reported in Ref.~\cite
  {martelli_hyperuniformity}}\BibitemShut {NoStop}%
\bibitem [{\citenamefont {Cahn}\ and\ \citenamefont
  {Hilliard}(1958)}]{cahn_hilliard}%
  \BibitemOpen
  \bibfield  {author} {\bibinfo {author} {\bibfnamefont {J.~W.}\ \bibnamefont
  {Cahn}}\ and\ \bibinfo {author} {\bibfnamefont {J.~E.}\ \bibnamefont
  {Hilliard}},\ }\href@noop {} {\bibfield  {journal} {\bibinfo  {journal} {J.
  Chem. Phys.}\ }\textbf {\bibinfo {volume} {28}},\ \bibinfo {pages} {258}
  (\bibinfo {year} {1958})}\BibitemShut {NoStop}%
\end{thebibliography}%

\end{document}